Nonlocal Modeling in High-Velocity Impact Failure of 6061-T6 Aluminum
F.R. Ahad[1], K. Enakoutsa[1†], K.N. Solanki[2*], and D.J. Bammann[3]

[1]Center for Advanced Vehicular Systems, Mississippi State University, 200 Research Boulevard, Mississippi State, MS 39762, USA
[2]School of Engineering of Matter, Transport and Energy, Arizona State University, Tempe, AZ 85287, USA
[3]Mechanical Engineering Department, Mississippi State University, Mississippi State, MS 39762

[†]enakoutsa@yahoo.fr (co-corresponding author)
[*]kiran.solanki@asu.edu (co-corresponding author)

## Abstract

In this paper, we present numerical simulations with local and nonlocal models under dynamic loading conditions. We show that for finite element (FE) computations of high-velocity, impact problems with softening material models will result in spurious post-bifurcation mesh dependency solutions. To alleviate numerical instability associated within the post-bifurcation regime, a characteristic length scale was added to the constitutive relations based on calibration of the series of different notch specimen tests. This work aims to assess the practical relevance of the modified model to yield mesh independent results in the numerical simulations of high-velocity impact problems. To this end, we consider the problem of a rigid projectile moving at a range of velocities between 89-107 m/s, colliding against a 6061-T6 Aluminum disk. A material model embedded with a characteristic length scale in the manner proposed by Pijaudier-Cabot and Bazant (1987), but in the context of concrete damage, was utilized to describe the damage response of the disk. The numerical result shows that the addition of a characteristic length scale to the constitutive model does eliminate the pathological mesh dependency and shows excellent agreements between numerical and experimental results. Furthermore, the application of a nonlocal model for higher strain rate behavior shows the ability of the model to address intense localized deformations, irreversible flow, softening, and final failure.

**Keywords:** Mesh dependence, Dynamic failure, Damage delocalization, BCJ model, Nonlocal modeling

## 1. Introduction

Recently, efforts to introduce a numerical length scale into continuum models have led to a resurgence of research in the area of generalized continua (e.g., see Dillon and Kratochvil (1970), Nunziato and Cowin (1979), Bammann and Aifantis (1982), Aifantis (1984), Bammann (1988), Brown et al. (1989), McDowell (1992), Zbib and Aifantis (1988), Fleck and Hutchinson (1993), Tvergaard and Needleman (1995), Gurtin (1996), Nix and Gao (1998), Ramaswamy and Aravas (1998), Gurtin (2000), Regueiro et al. (2002) and Solanki and Bammann (2010b)). This is partially due to the fact that the local theory treats a body as a "continuum" of particles or points, the only geometrical property being that of position. A closer look at materials reveals a complex microstructure of grains, subgrains, shear bands and other topological features of the distribution of mass that are not taken into account by classical local theories. If the observer is far enough removed from a grain, he will see only a point. But a theory that strips away all of the geometrical properties of a grain except for the position of its center of mass will certainly fail to explain the more complex aspects of its mechanical response. Furthermore, in codes modeling hyperbolic systems, the differential equations become elliptic in the post-bifurcation regime, but the boundary conditions are still prescribed for hyperbolic systems. Similarly, in static codes, elliptical systems transform into hyperbolic systems, but the associated boundary conditions are still prescribed for the original elliptical system. Moreover, most finite element (FE) simulations of high-velocity impact problems include classical constitutive models for inelastic deformation of heterogeneous materials (e.g.,

Vignjevic et al. (2012), Brunig and Driemeier (2007), Brunig and Steffen (2011), and Nicolas et al. (2012)). These models, however, insufficiently address at least two important issues: (a) the microstructure-size effects on the performance of materials under high-velocity loading conditions and (b) the numerical aspects associated with the post-bifurcation mesh dependency. The first problem has attracted the attention of Abu Al-Rub (2008), Abu Al-Rub and Kim (2008), Volger and Clayton (2008), Clayton (2005a), and Clayton (2005b). However, very few studies have addressed the post-bifurcation mesh dependency issues in high-velocity impact numerical simulations (e.g., see Abu Al-Rub and Kim (2009), where a characteristic length scale was added to a material model to predict converged results for ballistic limits of some heterogeneous material).

It is well-established that the addition of a characteristic length scale to constitutive models involving softening, either in the form of spatial gradients (Aifantis (1984), Bammann and Solanki (2010), and Solanki and Bammann (2010b)) or integral nonlocal terms (Pijaudier-Cabot and Bazant (1987), Leblond et al. (1994), Tvergaard and Needleman (1995), and Enakoutsa et al. (2007)), maintains the character of governing equations in the post-bifurcation regime of the material. Thus, the usual drawbacks (unlimited localization, bifurcation with infinite number of bifurcated branches, meaningless zero energy dissipated at failure when the FE size is infinitely small, and post-bifurcation pathological mesh dependency issues) encountered in practical applications of classical softening constitutive models no longer exist. Another motivation for the introduction of the length scale(s) into constitutive models is to capture more of the underlying physics (micro-structural features) of materials, as required by the design of new materials that must sustain extreme environmental conditions, while still using continuum models. An example of this type of modeling is Dillon and Kratochvil (1970)'s gradient model, which is rooted in the framework of Coleman-Gurtin continuum thermodynamics. The work of Aifantis (1995) also provides a review of strain gradient-type models, but it is not repeated here. Furthermore, the addition of length scale(s) to constitutive models stems from the need to solve boundary value problems at very small length scales where these models no longer apply. Fleck et al. (1994)'s seminal work on mechanical tests of small specimens has demonstrated that the yield strength starts to increase sharply with subsequent decrease in the specimen dimension when the specimen dimension reaches a critical value: the so-called "Hall-Petch effect." This dimension is smaller than the minimum scale of the applicability of classical constitutive models. Although molecular dynamics, dislocations dynamics, and discrete polycrystalline simulations may be helpful to explain the "Hall-Petch effect," these frameworks are too computationally expensive to track full micro-system level behaviors. However, they are useful to understand the material's intrinsic physical length scales, which manifest themselves in materials failure as a width of shear bands according to Johnson (1987), strain gradient concentrating stress around second particles in metal alloys as reported by Courtney (2000), and strain gradients in microscopic process zones at crack tips according to Hutchinson (2000). The addition of a length scale to the continuum theory using spatial gradients has predicted satisfactorily the Hall-Pecth effect (Fleck et al. (1994).

In this paper, such a characteristic length scale is added to a physically-motivated internal state variable material model, the BCJ[1] model (Bammann (1984), Bammann and Johnson (1984), Bammann (1985), Bammann and Aifantis (1987), Bammann and Johnson (1987a), Bammann and Aifantis (1987b), Bammann (1988), Bammann et al. (1990b), Bammann (1990a), Bammann et al. (1993), Bammann et al. (1995), Bammann et al. (1996), Tucker et al. (2010), Solanki et al. (2010a)) to overcome pathological post-bifurcation mesh dependency issues in numerical simulations of high-velocity impact problems. These numerical difficulties arise once the predicted material response in an element begins to soften; that is, for dynamic problems, the moment when the system of differential equations changes from hyperbolic to elliptic, and for static problems, the reverse, either leading to ill-posed problems because the boundary and initial conditions for one system of equations are not suitable for the other. The use of internal state variables allows high-fidelity reproduction of finite deformation and temperature histories, damage, and high rate phenomena that occur during the impact. The characteristic length scale is introduced in the BCJ model via Pijaudier-Cabot and Bazant (1987)'s nonlocal damage approach, which has historically and contemporaneously demonstrated its power to represent complex, experimentally-observed phenomena (Bazant and Ozbolt (1990) and Enakoutsa et al. (2007)). In the context of concrete damage, Pijaudier-Cabot and Bazant (1987) have proposed a formulation in which only the softening variable is nonlocal, while the strain, the stress, and other variables retain their local definition. Following this suggestion, a nonlocal evolution equation for the damage within an otherwise unmodified BCJ model is adopted in the present study. The damage internal state variable is expressed as the spatial convolution of a "local damage" and a Gaussian weighting function. The width of this function includes a characteristic length scale in the resulting nonlocal BCJ model; this length physically represents a microstructural distance connected to the mean spacing between neighboring microstructures and calibrated using series of different notch specimen tests. The physical justification of the convolution operation is that, unlike the other internal state variables in the BCJ model, the damage cannot be defined for volumes smaller than the microstructure spacing and therefore is an essentially nonlocal quantity. Hence, the nonlocal extension of the BCJ model remains physically-motivated, as its original local version.

---

[1]Bammann-Chiesa-Johnson

Pijaudier-Cabot and Bazant (1987)'s proposal has been successfully extended to creep (Saanouni et al. (1989), and ductile fracture, Leblond et al. (1994), Tvergaard and Needleman (1995), and Enakoutsa et al. (2007)). Tvergaard and Needleman (1995) checked that it indeed eliminates the mesh size effects; also, Enakoutsa et al. (2007) demonstrated that, with a minor modification, it reproduces very well the results of typical ductile rupture experiments.

The objective of this work is to critically assess the robustness of Pijaudier-Cabot and Bazant (1987)'s suggestion on the more complex problem of high-velocity impact damage through the ability of the method to predict a mesh-independent damage propagation during the impact. And also, to use series of notch specimen tests to motivate/calibrate required length scale parameters. The use of this nonlocal form is appropriate because high-velocity impact problems exhibit some softening due to the increase of damage. The layout of the rest of this paper is as follows: Section 2 summarizes the successive physical mechanisms responsible for failure during high-velocity impact of metals. This mechanism often occurs in two consecutive stages: formation of adiabatic shear bands, very quickly after the time of the contact, then failure by ductile rupture in the shear bands. Section 3 provides the constitutive equations for the original BCJ model and its nonlocal variant. Section 4 is devoted to the numerical application of the local and nonlocal BCJ models on two different problems. Here, all numerical simulations were performed using the LS-DYNA explicit FE code. The first problem was related to the failure of a series of notched-tensile specimens made of a 6061-T6 aluminum alloy to calibrate the nonlocal length scale parameter. The results show that experiments are well described by the modified nonlocal BCJ model. The second application consists of a rigid striker moving at very high speed impacting horizontally against a 6061-T6 aluminum alloy disk. Here also, the numerical results show that the modified BCJ model predictions compared very well with the experimental test results. This establishes the relevance of the nonlocal concept to deal with high rate damage phenomena.

## 2. Physics of high-velocity impact damage

The physical mechanisms responsible for failure during high-velocity impact initiate at the time of the contact, with local heating arising in the impact region as a result of intense plastic shear deformation. This heat is generated so quickly that the effects of conduction are negligible; hence, the process can be considered adiabatic (temperature rises at a rate proportional to the plastic work rate). As a result, thermal softening occurs in the impact region while the surrounding material continues to harden. When the load increases, the local region deforms more than the surrounding material and narrow adiabatic shear bands are formed. While these bands do not deteriorate the material as cracks do, they are precursors to fracture.

Typically, nucleation, growth, and coalescence and/or cracks may appear in the shear bands when the impact load exceeds the local material strength. The failure process of a high impact velocity problem will then occur in two stages: (1) local heating in the impact region leading to the development of adiabatic shear bands and (2) ductile rupture in the shear bands. The thermal softening involved in this rupture process is not only significant for high-velocity impact but also for low-velocity impact. In fact, thermal softening effects can be observed even for a strain rate of the order of $1.0 \ s^{-1}$ for a wide number of steels, including 304L stainless steel. Figure 1 shows earlier experimental and theoretical (using a local form of the BCJ model) work of Bammann et al. (1993), which provides evidence that thermal softening effects occur in torsion tests of thin-walled tubes of 304L stainless steel at room temperature for a strain rate equal to $10.0 \ s^{-1}$. The ISV-based BCJ material model captures the influence of load-path changes, strain rate, and temperature history on the macroscale constitutive (stress-strain) response and includes the effects of the adiabatic heating. This particular ISV plasticity-damage model has been previously verified and validated for a number of metal alloys through extensive simulations and physical testing, see Horstemeyer (2001). The local and nonlocal versions of this model are presented in the next section.

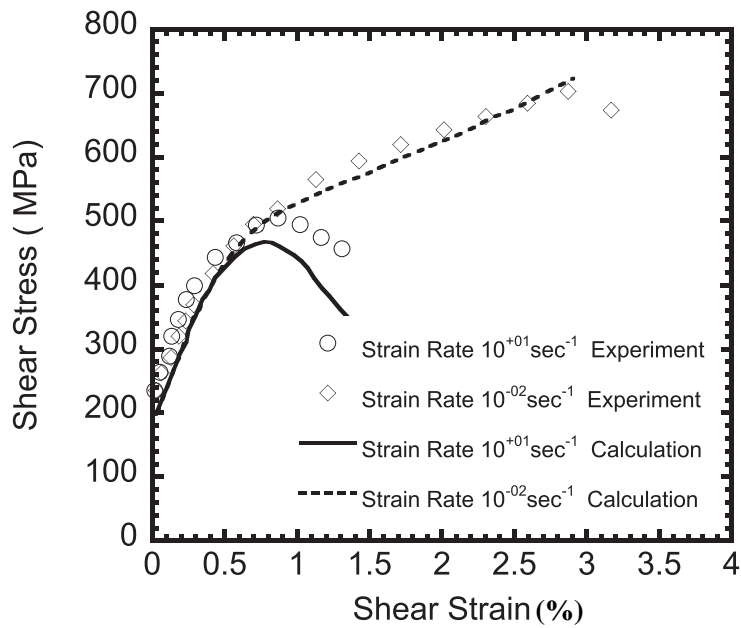

Figure 1: Comparison of finite element (FE) simulated torsional mechanical responses of 304L stainless steel thin-walled tubes at 20 °C for strain rates of 0.01 and 10.0 $s^{-1}$. Note the thermal softening effects. Calculations are shown by lines and data by markers. The curves are extracted from Bammann et al. (1993).

## 3. BCJ material model—original and nonlocal versions

### 3.1. The local model

The model presented here seeks to capture the combined effects of mechanical and thermal loading that materials may experience due to inelastic deformation and damage under dynamic loading conditions. Several versions of the BCJ model exist and are implemented in commercial and governmental FE codes (see Solanki et al. (2010a)). We are utilizing an older version of this model because it has been implemented in LS-DYNA, which allows us to take advantage of the nonlocal formulation that exists in that FE code. Nevertheless, several aspects of this model have been significantly modified in the last several years; see Solanki et al (2010a). The physics of the model have been modified to more accurately represent the dislocation microstructure that dictates the observed macroscopic response. This version of the model has also been cast in dimensionless form to both place tighter bounds on material parameters as well as reduce the variation of parameters (Solanki et al. (2009), Acar et al. (2010)) among the various materials (see also Marin et al. (2006)). In addition, the kinematics that follow have been modified to give consistent formulation between the kinematics and thermodynamics associated with the state variables representing the "continuum damage." Within this structure, the elastic moduli are automatically degraded with increasing damage, and the stress that is the major driving force in the inelastic flow is naturally concentrated with increasing damage. The particular form of these effects is determined by a single assumption of the form of the dependence of the damage deformation gradient upon the state variable representing the damage (see Solanki and Bammann (2010b) and Bammann and Solanki (2010)).

The kinematics employed in this early version of the model are an extension of the kinetics proposed by Davidson et al. (1977). Here we account for the deformation gradients associated with elastic, plastic, damage, and thermal expansion deformations. The constitutive models developed are written in rate form, but we begin by assuming a decomposition of the deformation gradient in the form

$$\boldsymbol{F} = \boldsymbol{F_e}\boldsymbol{F_d}\boldsymbol{F_p}\boldsymbol{F_\theta}. \tag{1}$$

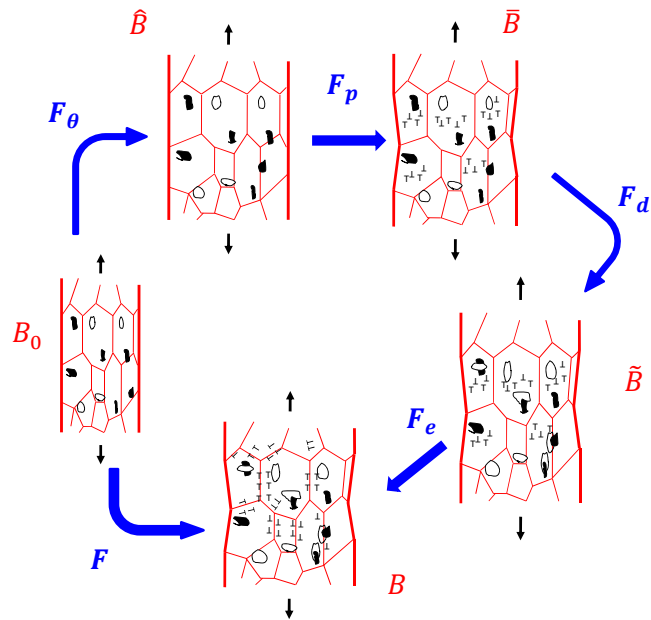

Figure 2: Multiplicative decomposition of the deformation gradient into the plastic, damage, thermal and elastic parts.

The elastic deformation gradient $\boldsymbol{F_e}$ describes the material point movement due to elastic motions and can be readily related to elastic unloading. The plastic deformation gradient $\boldsymbol{F_p}$ describes the material point movement due to the distortion caused by dislocation movement. The thermal deformation gradient $\boldsymbol{F_\theta}$ is due to thermal variation. The volumetric deformation gradient due to damage $\boldsymbol{F_d}$ describes the material point movement caused by dissipative volume changes of the material from nucleation, growth, and coalescences. In other words, if a void or defect is present, then enhanced dislocation nucleation, motion, and interaction would occur unlike if the void or defect were not present. As new internal free surfaces are created from the applied remote deformation, dislocations nucleate from the voids. One can think of this volume change related to dislocation nucleation to independently act upon void nucleation, growth, and coalescence. As such, each independent mechanism for damage can create internal free surfaces with dislocations independent of each other. The elastic deformation can be unloaded from the current configuration ($B$) to the intermediate configuration ($\tilde{B}$). The intermediate configuration ($\tilde{B}$) is a physically obtainable configuration by unloading elastically, while with a given damage, the configurations $\bar{B}, \hat{B},$ and $B_0$ are not obtainable (i.e., damage deformation is irrecoverable).

For most polycrystalline materials of interest, the thermal deformation gradient associated with thermal expansion is small; as such, the thermal portion of the deformation gradient can be given in terms of the linear coefficient of thermal expansion $\zeta$ and the temperature change $\Delta\theta$ as

$$\boldsymbol{F_\theta} = F_\theta \boldsymbol{I} = (1 + \zeta\Delta\theta)\boldsymbol{I}. \tag{2}$$

In the following section, we examine the effect of linearization of the elastic portion of the deformation on several kinematic measures. The polar decomposition of the elastic deformation gradient $\boldsymbol{F_e}$ is

$$\boldsymbol{F_e} = \boldsymbol{V_e R_e}, \tag{3}$$

where $\boldsymbol{R_e}$ and $\boldsymbol{V_e}$ are the elastic rotation and the elastic left stretch tensors, respectively. Typically in metals, the elastic strains are orders of magnitude less than plastic strains in well-developed plastic flow.

With the above deformation gradient components, we can define the following stretch tensors:

$$\boldsymbol{C} = \boldsymbol{F^T F}, \quad \boldsymbol{C_\theta} = \boldsymbol{F_\theta^T F_\theta}, \quad \widehat{\boldsymbol{C}}_p = \boldsymbol{F_p^T F_p}, \quad \overline{\boldsymbol{C}}_d = \boldsymbol{F_d^T F_d}, \quad \widetilde{\boldsymbol{C}}_e = \boldsymbol{F_e^T F_e}, \tag{4}$$

and the corresponding Lagrangian strain tensors:

$$\boldsymbol{E} = \tfrac{1}{2}(\boldsymbol{C} - \boldsymbol{I}), \quad \boldsymbol{E_\theta} = \tfrac{1}{2}(\boldsymbol{C_\theta} - \boldsymbol{I}), \quad \widehat{\boldsymbol{E}}_p = \tfrac{1}{2}(\widehat{\boldsymbol{C}}_p - \boldsymbol{I}), \quad \overline{\boldsymbol{E}}_d = \tfrac{1}{2}(\overline{\boldsymbol{C}}_d - \boldsymbol{I}), \quad \widetilde{\boldsymbol{E}}_e = \tfrac{1}{2}(\widetilde{\boldsymbol{C}}_e - \boldsymbol{I}). \tag{5}$$

Similar expressions for the damage strain have been discussed by Steinmann and Carol (1998), Hayakawa et al. (1998), Voyiadjis and Park (1999), Brünig (2001, 2003), and others.

Then the total strain is obtained by pulling all back (see Eq. 5) to the configuration $B_0$:

$$\boldsymbol{E} = \boldsymbol{E_e} + \boldsymbol{E_d} + \boldsymbol{E_p} + \boldsymbol{E_\theta}. \tag{6}$$

The fact that the total strain decomposes additively into the sum of elastic, damage, plastic, and thermal parts is true with respect to every configuration as long as each strain tensor has been properly "pushed forward" or "pulled back" between configurations.

The velocity gradient associated with the deformation gradient in the current configuration $B$, $\boldsymbol{l} = \dot{\boldsymbol{F}}\boldsymbol{F^{-1}}$ is separated into an elastic, plastic, thermal, and a volumetric part, and is given by

$$\boldsymbol{l} = \boldsymbol{l_e} + \boldsymbol{l_d} + \boldsymbol{l_p} + \boldsymbol{l_\theta}, \ \boldsymbol{l_e} = \dot{\boldsymbol{F}}_e \boldsymbol{F_e^{-1}}, \boldsymbol{l_d} = \boldsymbol{F_e}\dot{\boldsymbol{F}}_d \boldsymbol{F_d^{-1}} \boldsymbol{F_e^{-1}}, \boldsymbol{l_p} = \boldsymbol{F_e}\boldsymbol{F_d}\dot{\boldsymbol{F}}_p \boldsymbol{F_p^{-1}} \boldsymbol{F_d^{-1}} \boldsymbol{F_e^{-1}}, \text{ and}$$
$$\boldsymbol{l_\theta} = \boldsymbol{F_e}\boldsymbol{F_d}\boldsymbol{F_p}\dot{\boldsymbol{F}}_\theta \boldsymbol{F_\theta^{-1}} \boldsymbol{F_p^{-1}} \boldsymbol{F_d^{-1}} \boldsymbol{F_e^{-1}} \ . \tag{7}$$

As with the strain tensor, a similar additive equation holds for the velocity gradient with respect to every configuration. By pulling back the above equation through $\boldsymbol{F_e^{-1}}$, the velocity gradient in the intermediate configuration $\bar{B}$ results in

$$\bar{\boldsymbol{l}} = \bar{\boldsymbol{l}}_e + \bar{\boldsymbol{l}}_d + \bar{\boldsymbol{l}}_p + \bar{\boldsymbol{l}}_\theta, \ \bar{\boldsymbol{l}}_e = \boldsymbol{F_e^{-1}}\dot{\boldsymbol{F}}_e, \ \bar{\boldsymbol{l}}_d = \dot{\boldsymbol{F}}_d \boldsymbol{F_d^{-1}}, \bar{\boldsymbol{l}}_p = \boldsymbol{F_d}\dot{\boldsymbol{F}}_p \boldsymbol{F_p^{-1}} \boldsymbol{F_d^{-1}}, and \ \bar{\boldsymbol{l}}_\theta = \boldsymbol{F_d}\boldsymbol{F_p}\dot{\boldsymbol{F}}_\theta \boldsymbol{F_\theta^{-1}} \boldsymbol{F_p^{-1}} \boldsymbol{F_d^{-1}} \tag{8}$$

Note that we can decompose any velocity gradient into skew and symmetric parts in any configuration. For example, in the current configuration $B$,

$$\boldsymbol{l} = \boldsymbol{d} + \boldsymbol{w}, \ \boldsymbol{d} = sym(\boldsymbol{l}) = \tfrac{1}{2}(\boldsymbol{l} + \boldsymbol{l}^T), \quad \text{and} \quad \boldsymbol{w} = skew(\boldsymbol{l}) = \tfrac{1}{2}(\boldsymbol{l} - \boldsymbol{l}^T); \tag{9}$$

and in the intermediate configuration $\widetilde{B}$,

$$\tilde{\boldsymbol{l}} = \tilde{\boldsymbol{d}} + \tilde{\boldsymbol{w}}, \ \tilde{\boldsymbol{d}} = sym(\tilde{\boldsymbol{l}}) = \tfrac{1}{2}(\tilde{\boldsymbol{l}} + \tilde{\boldsymbol{l}}^T), \quad \text{and} \quad \tilde{\boldsymbol{w}} = skew(\tilde{\boldsymbol{l}}) = \tfrac{1}{2}(\tilde{\boldsymbol{l}} - \tilde{\boldsymbol{l}}^T). \tag{10}$$

With the derived expressions above, we can now map the velocity gradient and stress tensor between different configurations as shown in Table 1.

Table 1 Mapping of stress tensor and velocity gradient between respective configurations

| Configuration | Stress | Velocity gradient | Density |
|---|---|---|---|
| $B$ | $\boldsymbol{\sigma}$ | $\boldsymbol{l} = \dot{F}F^{-1}$ | $\rho$ |
| $\widetilde{B}$ | $\widetilde{S} = J_e F_e^{-1} \boldsymbol{\sigma} F_e^{-T}$ | $\tilde{l} = F_e^{-1} l F_e$ | $\tilde{\rho} = J_e \rho$ |
| $\overline{B}$ | $\overline{S} = J_d F_d^{-1} \widetilde{S} F_d^{-T}$ | $\bar{l} = F_d^{-1} \tilde{l} F_d$ | $\bar{\rho} = J_d \tilde{\rho}$ |
| $\hat{B}$ | $\widehat{S} = J_p F_p^{-1} \overline{S} F_p^{-T}$ | $\hat{l} = F_p^{-1} \bar{l} F_p$ | $\hat{\rho} = J_p \bar{\rho}$ |
| $B_0$ | $S = J_\theta F_\theta^{-1} \widehat{S} F_\theta^{-T}$ | $L = F_\theta^{-1} \hat{l} F_\theta$ | $\rho_0 = J_\theta \hat{\rho}$ |

$\rho$ is the density in the current configuration, $\boldsymbol{\sigma}$ is the Cauchy stress in the current configuration, $\widetilde{S}$ is the second Piola-Kirchhoff stress in the intermediate configuration, $\overline{S}$ is the second Piola-Kirchhoff stress in the $\overline{B}$ configuration, $\widehat{S}$ is the second Piola-Kirchhoff stress in the $\hat{B}$ configuration, and $S$ is the second Piola-Kirchhoff stress in the reference configuration.

The material time derivative of the strain tensors $(\widetilde{E}, \widetilde{E}_p, \widetilde{E}_e, \widetilde{E}_\theta)$ living in the intermediate configuration $\tilde{B}$ is

$$\dot{\widetilde{E}} = F_e^T dF_e - (\bar{l}^T - \bar{l}_e^T)\widetilde{E} - \widetilde{E}(\bar{l} - \bar{l}_e), \tag{11}$$

where

$$F_e^T dF_e = F_e^T d_e F_e + F_e^T d_d F_e + F_e^T d_p F_e + F_e^T d_\theta F_e$$
$$F_e^T d_p F_e = \tilde{d}_p + \bar{l}_p^T \widetilde{E}_e + \widetilde{E}_e \bar{l}_p$$
$$F_e^T d_\theta F_e = \tilde{d}_\theta + \bar{l}_\theta^T \widetilde{E}_e + \widetilde{E}_e \bar{l}_\theta$$
$$F_e^T d_d F_e = \tilde{d}_d + \bar{l}_d^T \widetilde{E}_e + \widetilde{E}_e \bar{l}_d$$
$$F_e^T d_e F_e = \dot{\widetilde{E}}_e$$
$$F_d^T \tilde{d}_p F_d = \bar{d}_p + \bar{l}_p^T \overline{E}_d + \overline{E}_d \bar{l}_p$$

$$\tilde{d}_p = \frac{\bar{l}_p^T + \bar{l}_p}{2}.$$

Similarly, we can relate the time derivative of the elastic, the plastic, the thermal, and the volumetric strain tensors to the elastic, the plastic, the thermal, and the volumetric rate of deformation tensors as follows:

$$\dot{\widetilde{E}}_p = \tilde{d}_p - (\bar{l}_p^T + \bar{l}_d^T)\widetilde{E}_p - \widetilde{E}_p(\bar{l}_p + \bar{l}_d) - \bar{l}_p^T \widetilde{E}_d - \widetilde{E}_d \tilde{l}_p \tag{12}$$

$$\dot{\widetilde{E}}_\theta = \widetilde{d}_\theta - (l^T - l_e^T)\widetilde{E}_\theta - \widetilde{E}_\theta(\bar{l} - \bar{l}_e) - l_\theta^T(\widetilde{E}_p + \widetilde{E}_d) - (\widetilde{E}_p + \widetilde{E}_d)\bar{l}_\theta \tag{13}$$

$$\dot{\widetilde{E}}_d = \widetilde{d}_d - \bar{l}_d^T\widetilde{E}_d - \widetilde{E}_d\bar{l}_d \tag{14}$$

$$\dot{\widetilde{E}}_e = F_e^T d_e F_e = F_e^T dF_e - \widetilde{d}_p - \widetilde{d}_\theta - \widetilde{d}_d - (l^T - l_e^T)\widetilde{E}_e - \widetilde{E}_e(\bar{l} - \bar{l}_e). \tag{15}$$

To complete this system of equations, it is necessary to write expressions for $\widetilde{d}_p, \widetilde{d}_\theta,$ and $\widetilde{d}_d$ in Eq. (15) as well as $\bar{l} - \bar{l}_e$, which necessitates the additional representation of $\widetilde{w}_p, \widetilde{w}_\theta,$ and $\widetilde{w}_d$. These equations have been discussed in Solanki et al. (2011). The rotational parts are necessary to specify the convective parts of the derivatives in Eq. (15) that, in conjunction with the expression of companion stretches, fixes or specifies the position of each configuration at any given instant as can been seen by properly taking the material derivative of these kinematic quantities by pulling back to the current configuration, taking the time derivative and pushing forward. The required rotation and rotational rate naturally appear to properly render the derivatives objective.

In present theory, we can define damage $\phi$ as the ratio of the change in volume of an element in the elastically unloaded state ($\bar{B}$) from its volume in the initial reference state to its volume in the elastically unloaded state:

$$\phi = \frac{V_v}{V_{\bar{B}}}, \tag{16}$$

where $V_{\bar{B}}$ is the total volume in the intermediate configurations and $V_v$ is the added volume due to voids. The change in volume from the reference configuration $B_0$ to the intermediate configuration $\widetilde{B}$ is

$$V_{\bar{B}} = V_{B_0} + V_v, \tag{17}$$

where $V_{B_0}$ is the initial volume in the reference configuration. Using Eq. 16 and Eq. 17, we can show that

$$V_{B_0} = V_{\bar{B}}(1 - \phi), \tag{18}$$

where now the Jacobian is determined by the damage parameter $\phi$:

$$J_d = det F_d = \frac{V_{\bar{B}}}{V_{B_0}} = \frac{1}{(1-\phi)}. \tag{19}$$

From this definition, we get

$$F_d = \frac{1}{(1-\phi)^{1/3}}I \tag{20}$$

$$\bar{l}_d = \dot{F}_d F_d^{-1} = \frac{\dot{\phi}}{3(1-\phi)}I, \tag{21}$$

where $I$ is the identity tensor.

For the evolution equation of isotropic damage, we consider the non-associative form of Cocks and Ashby (1980) as shown in Eq. 22:

$$\dot{\phi} = |\boldsymbol{d_p}| \sinh\left[\frac{2(2m-1)}{2m+1} \frac{p}{\{\|\boldsymbol{\sigma'} - \boldsymbol{\alpha}\|\}}\right] \left(\frac{1 - [1 - \phi]^{m+1}}{[1 - \phi]^m}\right). \tag{22}$$

Let us now define a measure of the stress $\tilde{\boldsymbol{S}}$ in the intermediate configuration expressed in terms of the Cauchy stress as

$$\tilde{\boldsymbol{S}} = J_e \boldsymbol{F_e^{-1}} \boldsymbol{\sigma} \boldsymbol{F_e^{-T}}, \tag{23}$$

where $J_e = det \boldsymbol{F_e}$. The constitutive equation relating the stress to the elastic strain in this configuration has the usual linear form:

$$\tilde{\boldsymbol{S}} = \left[2\mu(\theta, \phi)\, \tilde{\boldsymbol{E}}'_e + K(\theta, \phi)\, tr(\tilde{\boldsymbol{E}}_e)\boldsymbol{I}\right], \tag{24}$$

where $K(\theta, \phi)$ and $\mu(\theta, \phi)$ are the damage and temperature dependent bulk and shear moduli, respectively, and $\tilde{\boldsymbol{E}}'_e$ is the deviatoric part of $\tilde{\boldsymbol{E}}_e$. In this implementation of the BCJ model, the temperature dependence of the elastic moduli is neglected and the damage dependence of the elastic moduli is the familiar form valid for the dilute concentration of voids:

$$\mu(\phi) = \mu_0(1 - \phi) \text{ and } K(\phi) = K_0(1 - \phi), \tag{25}$$

where $K_0$ and $\mu_0$ are the bulk and shear moduli in the undamaged material. Taking the time derivative of Eq. (24),[2] the constitutive relation can be represented in terms of a convective derivative of the Cauchy stress in the current configuration, which involves the skew-symmetric part of the elastic velocity gradient tensor:

$$\dot{\boldsymbol{\sigma}}' - \boldsymbol{w_e}\boldsymbol{\sigma}' - \boldsymbol{\sigma}'\boldsymbol{w_e^T} = \lambda(\phi)\, tr(\boldsymbol{d_e})\boldsymbol{I} + 2\mu(\phi)\boldsymbol{d_e}' - \frac{4}{3}\boldsymbol{\sigma}'\frac{\dot{\phi}}{(1-\phi)}, \tag{26}$$

where the elastic stretching and spin are

$$\boldsymbol{d_e} = \boldsymbol{d} - \boldsymbol{d_d} - \boldsymbol{d_p} - \boldsymbol{d_\theta} \text{ and } \boldsymbol{w_e} = \boldsymbol{w} - \boldsymbol{w_p}. \tag{27}$$

The plastic stretching $\boldsymbol{d_p}$ is assumed to occur in the direction of the net deviatoric stress, which is the stress minus/plus any internal stresses. This results in the expression of the form

$$\boldsymbol{d_p} = \dot{\gamma}\left(\frac{\boldsymbol{\sigma'} - \boldsymbol{\alpha}}{\|\boldsymbol{\sigma'} - \boldsymbol{\alpha}\|}\right), \tag{28}$$

---

[2]This assumption is unnecessary when employing the approach developed in Bammann and Solanki (2010) and Solanki and Bammann (2010b).

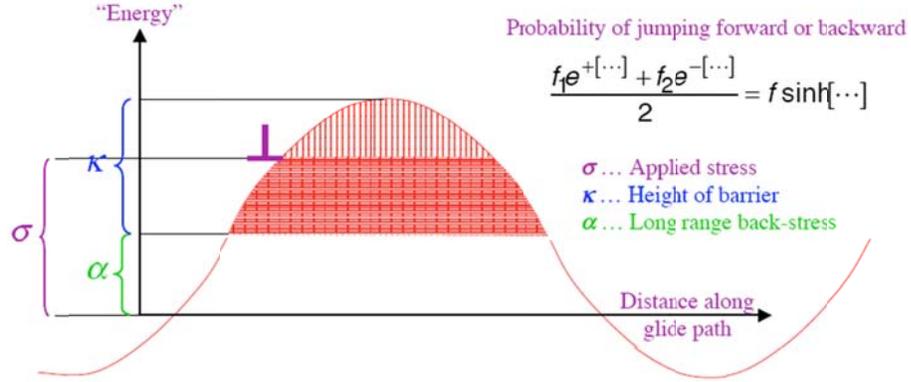

Figure 3: Formulation of flow rule motivated by energetic required for overcoming obstacles (Bammann, 1988).

where $\dot{\gamma}$ is the magnitude of the plastic flow. The evolution equation $\dot{\gamma}$ is formulated in context of the energy required for overcoming obstacles (see Figure 3) and also includes texture effects (Bammann (1988)). Therefore, $\dot{\gamma}$ has a strong nonlinear dependence upon the activation energy of the barriers. We therefore assume $\dot{\gamma}$ of the form

$$\dot{\gamma} = f(\theta)\left[\sinh\left(\frac{-\Delta G\|\xi\|}{V(\theta)}\right)\right], \tag{29}$$

where $V(\theta) = C_1 e^{(-C_2/\theta)}$ and $f(\theta) = C_5 e^{(-C_6/\theta)}$ are related to yielding with Arrhenius-type temperature dependence and $\Delta G\|\xi\|$ is the activation energy dislocations to overcome local obstacles, which depends upon the local net stress. The hyperbolic sine is a result of assuming equal probability of forward and backward jumps over obstacles. If we assume that the main source of obstacles that must be overcome are the statistically stored dislocations, then one choice for the activation energy representing square profiles is

$$\Delta G\|\xi\| = \frac{\|\sigma' - \alpha\| - (\kappa + Y(\theta))(1-\phi)}{(1-\phi)}, \tag{30}$$

where $\|\boldsymbol{\sigma}' - \boldsymbol{\alpha}\|$ is the damage-concentrated stress minus "back stress" in the current configuration. The specific form of the damage concentration results from the specific choice made previously for the inverse metric dependence upon the damage, and $\kappa$ represents an averaged internal strength of the material and is the thermodynamic conjugate to the elastic strain associated with statistically stored dislocations introduced previously. Similarly, $\alpha$ is the thermodynamic conjugate to the incompatible lattice curvature due to presence of geometrically necessary dislocations (GNDs) at grain boundaries and around second-phase particles. Motivated by these concepts, the expression for the plastic stretching is chosen in the form

$$\boldsymbol{d_p} = f(\theta)\left[\sinh\left(\frac{\|\boldsymbol{\sigma}' - \boldsymbol{\alpha}\| - (\kappa + Y(\theta))(1-\phi)}{V(\theta)(1-\phi)}\right)\right]\left(\frac{\boldsymbol{\sigma}' - \boldsymbol{\alpha}}{\|\boldsymbol{\sigma}' - \boldsymbol{\alpha}\|}\right). \tag{31}$$

The functions $Y(\theta) = C_3 e^{(-C_4/\theta)}$ is related to yielding with Arrhenius-type temperature dependence. where $C_1$ through $C_6$ are the yield stress related material parameters that are obtained from isothermal compression tests with variations in temperature and strain rate, see Appendix A.

We view plastic deformation as the motion of dislocations and the state of the material as a freeze-frame of the deformed state, which is represented by the elastic lattice deformation due to the presence of dislocations and due to the external loading (see Figure 4). We relate this strain-like internal variable $\tilde{\varepsilon}_s$ to the density of statistically-stored dislocations $\tilde{\rho}_s$. In an increment of strain, dislocations are stored inversely proportional to the mean free path, which in a Taylor lattice is inversely proportional to the square root of the dislocation density. Dislocations are annihilated or "recover" due to cross slip or climb in a manner proportional to the dislocation density. Following the Taylor assumption (Bammann, 2001), the lattice deformation due to the presence of statistically stored dislocations $\tilde{\varepsilon}_s$ in the intermediate configurations can be defined as

$$\tilde{\varepsilon}_s = b\sqrt{\tilde{\rho}_s} \, , \tag{32}$$

where $b$ is the magnitude of Burger's vector (see Figure 4).

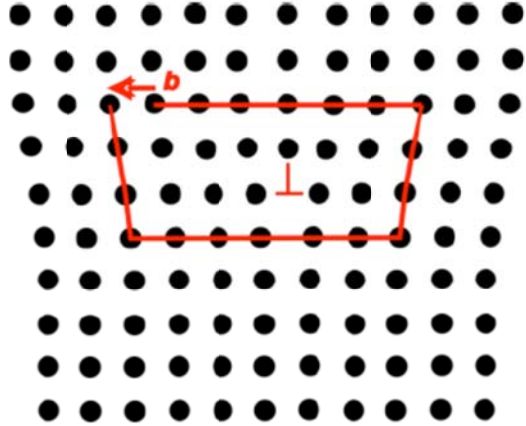

Figure 4: Statistically stored dislocations (SSD) provide observed hardening.

Mecking and Kocks proposed that during an increment of plastic strain, the density of the statistically stored dislocations is stored inversely proportional to the mean free path and recovered proportionally to the density of the dislocations

$$\frac{\partial \tilde{\rho}_s}{\partial \tilde{\varepsilon}_s} = \frac{c_7}{l} - C_8 \tilde{\rho}_s \, , \tag{33}$$

where $l$ is the mean free path of the dislocations, and $C_7$ and $C_8$ are constants. Motivated by this, we choose a hardening law of the form

$$\dot{\kappa} = (H(\theta) - R_d(\theta)\kappa^2)\|\boldsymbol{d_p}\| - R_s(\theta)\kappa^2. \tag{34}$$

The dynamic recovery $R_d(\theta)$ is motivated from dislocation cross slip that operates on the same time scale as dislocation glide. For this reason, no additional rate dependence results from this recovery term. The thermal recovery term $R_s(\theta)$ is related to the diffusional process of vacancy-assisted climb. Because this process operates on a much slower time scale, strong rate dependence is predicted at higher temperatures where this term becomes dominant. For the stress-like internal state variables work-conjugate $\boldsymbol{\alpha}$ to GND, we will follow the form chosen by Armstrong et al. (1966), who postulated that the "back stress" hardens in the direction of plastic flow and recovers in the direction of $\boldsymbol{\alpha}$. To incorporate effects of direction changes, we include the magnitude $\boldsymbol{\alpha}$ in the recovery. This type of evolution is proposed based upon dislocation pileups occurring at grain boundaries or particles during loading, and "running" in the opposite direction during load reversal. This results in an apparent softening under the load reversal.

Since dislocation pileups are also geometrically necessary dislocations, this is an adequate evolution equation as a first approximation:

$$\dot{\boldsymbol{\alpha}} - \boldsymbol{w_e}\boldsymbol{\alpha} - \boldsymbol{\alpha}\boldsymbol{w_e^T} = h(\theta)\,\boldsymbol{d_p} - \left(r_s(\theta) + r_d(\theta)\left\|\boldsymbol{d_p}\right\|\right)\|\boldsymbol{\alpha}\|\boldsymbol{\alpha}. \tag{35}$$

The scalar $\kappa$ is an isotropic hardening variable that predicts no change in flow stress upon reverse loading. This variable captures long transients and is responsible for the prediction of continued hardening at large strains. Once steady state has been reached under constant conditions, this variable is not affected by a change in loading, though it is still affected by change in temperature. In Equations (34) and (35), $r_d(\theta)$ and $R_d(\theta)$ are scalar functions of temperature that describe dynamic recovery whereas $r_s(\theta)$ and $R_s(\theta)$ are scalar functions that describe thermal (static) recovery with $h(\theta)$ and $H(\theta)$ representing the anisotropic and isotropic hardening modulus, respectively. These functions are calculated as

$$r_d(\theta) = C_7 e^{(-C_8/\theta)} \tag{35a}$$

$$R_d(\theta) = C_{13} e^{(-C_{14}/\theta)} \tag{35b}$$

$$r_s(\theta) = C_{11} e^{\left(-C_{12}/\theta\right)} \tag{35c}$$

$$R_s(\theta) = C_{17} e^{\left(-C_{18}/\theta\right)} \tag{35d}$$

$$h(\theta) = C_9 e^{\left(-C_8/\theta\right)} - C_{10}\theta \tag{35e}$$

$$H(\theta) = C_{15} e^{\left(-C_8/\theta\right)} - C_{16}\theta \tag{35f}$$

where $C_7$ through $C_{12}$ are the material plasticity parameters related to kinematic hardening and recovery terms, and $C_{13}$ through $C_{18}$ are the material plasticity parameters related to isotropic hardening and recovery terms, respectively, see Appendix A. Constants $C_1$ through $C_{18}$ are found from macroscale experiments (i.e., tension, compression, and shear tests) at different temperatures and strain rates, see Appendix A.

The last equation to complete the description of the model is one that computes the temperature change during high strain-rate deformations, such as those encountered in high-rate impact loadings. For these problems, a non-conducting (adiabatic) temperature change, following the assumption that 90% of the plastic work is dissipated as heat, is assumed. Taylor and Quinney (1934) were the first to measure the energy dissipation from mechanical work as being between 5−50% of the total work for various materials and strain-rate levels. The rate of the change of the temperature is assumed to follow the equation

$$\dot{\theta} = \frac{0.9}{\rho C_v}\left[\boldsymbol{\sigma} : \boldsymbol{d_p}\right], \tag{36}$$

where $\rho$ and $C_v$ represent the material density and a specific heat coefficient, respectively. The empirical assumption in Eq. (36) has permitted a non-isothermal solution by FE that is not fully coupled with the energy balance equation (Bammann et al. (1993)). Note that the temperature rise induces a profound effect on the constitutive behavior of the material. For example, the temperature increase leads to thermal softening (adiabatic shear bands), and as a result, shear instabilities may arise.

This particular ISV plasticity-damage model has been previously verified and validated for a number of metal alloys through extensive simulations and physical testing. For example, Horstemeyer (2001) successfully employed the ISV plasticity-damage constitutive model for structural analysis of components made of A356 cast aluminum alloy. Through appropriate multiscale modeling, his team was able to correctly predict the failure mode of an automobile control arm under multiple loading conditions. We expect a substantial development of this model in the future, provided that the unlimited localization phenomenon and the resulting post-bifurcation mesh dependency it predicts in problems involving localization is solved. Indeed, the FE solution of a problem involving strain or damage localization (e.g., shear bands) using material models including softening, such as the BCJ model, becomes meaningless when one element begins to soften, and all the deformation has the tendency to concentrate within a zone approaching zero measure with decreasing element size. Because the energy is proportional to the deformation, the predicted dissipated energy gradually reduces with decreasing element size; when the FE size tends to zero, the dissipation energy also goes to zero. The explanation for this phenomenon is that the elements outside the localization zone unload after the occurrence of localization in one element. As a consequence, the energy dissipation decreases with decreasing size of the softening element. These difficulties disappear when a nonlocal concept is introduced in the constitutive model. The following section presents a technique to introduce the nonlocal concept into the BCJ model.

## 3.2. Nonlocal BCJ model

Many materials, when at room or elevated temperature, undergo softening, i.e., stress decreases while the strain increases. Often this property is related to a gradual increase of the damage, which manifests as micro-crack growth and void formation. The increase of damage is responsible for material stiffness reduction, which eventually leads to softening that is well-known to be at the origin of the material instability. Unfortunately, the presence of softening in material models produces undesirable unlimited localization of strain, resulting in meaningless zero dissipation energy during crack propagation and pathological mesh sensitivity in FE computations. These difficulties resolve when a nonlocal concept is adopted. Nonlocal theories were first introduced by Kröner (1967), Krumhans (1968) and Kunin (1968), and developed later by Eringen (1972), Eringen and Edelen (1972), Eringen and Ari (1983), and Edelen (1969) in the context of elasticity. Eringen (1972)'s nonlocal formulation involves weighted spatial averages of quantities. Pijaudier-Cabot and Bazant (1987) extended this nonlocal concept to strain softening materials, mainly concrete and rocks; Saanouni et al. (1989) and Leblond et al. (1994) applied this concept to creep and ductile fracture problems, respectively. Here we follow Pijaudier-Cabot and Bazant (1987)'s suggestion to introduce the nonlocal concept into the BCJ material model to address the problems related to the presence of softening in the model.

In the local BCJ model, softening may arise from two mechanisms: a gradual increase of the damage (under isothermal conditions) or a temperature rise (in adiabatic conditions) followed by an increase in damage. However, we do not follow Kane et al. (2009)'s suggestion to delocalize the two variables causing softening in the constitutive relation. In fact, review of the model's constitutive equations provided in the previous section reveals that temperature and damage parameters are indeed related toward their sequential governing equations of softening in adiabatic conditions. It appears quite appealing from a physical point of view to introduce a nonlocal concept into the governing equations through the damage equation. Indeed, just like in heterogeneous materials, defining the damage requires consideration of the "elementary" volumes of size greater than the void spacing[3]; therefore, the damage is essentially a nonlocal quantity. The equation for the nonlocal variable is assumed to have the form of a Gauss distribution function the width of which introduces a mathematical length scale. This equation reads

---

[3] The Cocks and Ashby (1980) void growth model is based on a cylinder containing a spherical void.

$$\phi(X) = \frac{1}{B(X)} \int_\Omega \phi^{loc}(Y) A(X-Y) \, d\Omega_y. \tag{37}$$

In Eq. (37), $\Omega$ denotes the volume of the body studied, $X$ and $Y$ the vector point coordinates, and A the Gauss distribution function which is defined as

$$A(X) = exp\left(-\frac{\|X\|^2}{L^2}\right), \tag{38}$$

In Eq. (38), the term $L$ is the characteristic length scale, $\|X\|$ represents the magnitude of the vector $X$, while the function B(x) is defined as

$$B(X) = \int_\Omega A(X-Y) \, d\Omega_y. \tag{39}$$

The "local damage" $\phi^{loc}$ is deduced from Eq. (37). The function A is infinitely differentiable; it does not include any Dirac's δ-distribution at the point 0. This ensures that the function $\phi$ is entirely nonlocal. Other properties of the function A include isotropy and normalization. As a consequence, the function $\phi$ must be equal to $\phi^{loc}$ if the latter variable is spatially uniform. In the absence of the normalization factor 1/B, this would not be the case near the boundary of $\Omega$. The presence of the normalization term allows for the coincidence everywhere. Another weighing function,

$$A(X) = \left(\frac{1}{1+\left(\frac{\|X\|}{L}\right)^p}\right)^q, \tag{40}$$

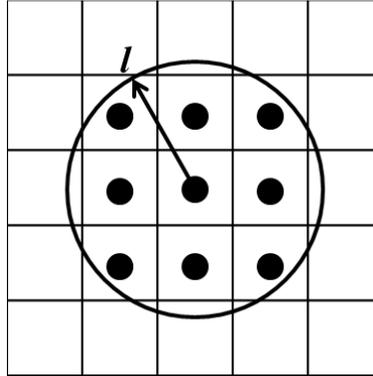

Figure 5: Idealization of the nonlocal interaction zone

where $L$ is the material characteristic length scale which will be calibrated through series of tensile notch tests, can be used instead of Eq. (38). In Eq. (40), the coefficients p and q are chosen to be constant and equal to p=8 and q=2, so that in the interaction zone determined by the length scale, i.e., Eq. (38), accounts for strong non-locality effects (see Figure 5). The constitutive equations of the nonlocal BCJ model consist of all the equations presented in Sub-section 3.1 except the evolution equation for the damage (Eq. (22)), which is replaced by Eq. (37) where, again, the "local damage" is calculated from the "local damage rate" found in Eq. (22). The nonlocal (integral) expression of the equation of the damage is critical to understanding why unlimited localization of strain and damage is prevented with a nonlocal BCJ model in problems involving localization. Indeed, Eq. (37) given by an integral expression is continuous, and the plastic dissipation energy is proportional to the nonlocal rate of the damage, as in Bazant and Pijaudier-Cabot (1988). Hence, the dissipation energy is spatially continuous and cannot localize into zones of vanishing width. Interestingly, the continuity of the dissipation energy yields the continuity of the time-derivative of the temperature because, according to Eq. (36), the latter is proportional to the dissipation energy. Therefore, the time-derivative of the temperature is also spatially continuous. Consequently, the concentration of any adiabatic shear band in the region of zero width is prohibited in nonlocal BCJ media. Another direct consequence is that the jump of the temperature rate toward any surface of discontinuity is zero, which is critical to show that no spatial discontinuity can occur in the velocity gradient despite the softening arising from a gradual increase of the damage. The theoretical explanations of why unlimited localization is prevented within the nonlocal BCJ model are widely commented in a previous paper by the authors (Enakoutsa et al. (2012)). The authors also show in the same paper that the tangent stiffness tensor in the nonlocal BCJ medium reduces to the elasticity tensor, which is positive definite. Hence, all the issues associated with the loss of positive-definiteness of the stiffness matrix (which include pathological mesh-size dependency, non-uniqueness of the solution for the governing differential equations at bifurcations, zero energy dissipation at failure when the FE size is very small) at the onset of softening, as encountered in the local BCJ medium (and reported in other elasto-plastic model by Bazant and Belytschko (1985), Neilson and Schreyer (1993), and Rudinicki and Rice (1975)), are automatically resolved. It follows that the FE results predicted by nonlocal models are objective. The works of Peerling et al. (1998, 1996), and Sluys and de Borst (1992) yield similar conclusions.

### 3.3. Significance of the characteristic length for the nonlocal BCJ model

The nonlocal damage evolution (see Eq. (37)) includes a characteristic length scale $L$ aimed at solving the pathological post-bifurcation mesh-dependency issues in numerical simulations involving the BCJ model, more specifically in high-velocity impact problems of interest in this paper. In theory, this parameter is interpreted as the mean half-spacing between neighboring microstructures, but in real materials the microstructure distribution is always random so that it is difficult to determine its precise value through microstructural observations. However, Tvergaard and Needleman (1997)'s work showed that for porous ductile materials, the characteristic length scales, with either the void spacing or the mean void radius, depend on whether the mechanism leading to the final failure is a coalescence of voids or the formation of shear bands. For concrete materials, Bazant and Pijaudier-Cabot (1989) have demonstrated that the characteristic length is proportional to the maximum aggregate size. Also, Abu Al-Rub (2004) used a dislocation-based argument to relate the material length scale to the mean free-path for dislocation motion and showed that this length is a function of the material microstructure features, such as mean grain size in polycrystalline materials or the mean particle size in particle-reinforced composites. Furthermore, Voyiadjis et al. (2008)'s constitutive relations suggested a non-constant characteristic length scale, i.e., evolving as a function of different mechanisms associated to softening. Recently, Enakoutsa et al. (2007) and Enakoutsa and Leblond (2009) characterized this length scale for A508 Cl.3 steel described by the nonlocal extensions of Gurson (1977)'s ductile fracture model by comparing model predictions and experimental results for typical laboratory-scale ductile fracture tests. By trial and error, they found this length scale to be of the order of 0.55 mm. However, it is not clear whether the characteristic length scale is an intrinsic material parameter or simply a parameter that allows the numerical computations to yield objective results. We shall assume that it is an adjustable parameter, similar to the minimum element size, as suggested Rousselier (1981); this minimum element size must be chosen in more typical numerical simulations based on the BCJ model. However, our objective here is to calibrate such a length scale using a series of tensile notch test specimens with varying notch radii. In a uniform material, the notch geometry induces a smooth-stress triaxiality field near the center of the specimen and the stress triaxiality is the primary driving factor for damage in porous materials, post-bifurcation issues arise. Furthermore, the only other constraint of this assumption is that the characteristic length scale must be greater than (or at least equal to) the element size in regions where the impact damage may occur. It is only at this condition that the regularization by nonlocal averaging is viable.

## 4. Numerical applications: High-velocity impact simulations

### 4.1. Mesh sensitivity in high-velocity impact simulations: Generalities

The need for new materials in the design of protective systems for civilian or military applications requires these enhanced material perform under high-velocity impact-loading conditions. To that end, the existing design tools for such materials must be improved to describe material behavior with high fidelity, especially after impact-damage initiation. However, the majority of material models currently used in numerical simulations of high-velocity impact damage fail to represent the material's post-bifurcation regime because they do not include any characteristic length scales. In the few models that do (e.g., the model by Needleman (1988) works for rate-sensitive material models), the (physical) characteristic length scale is so small that any meaningful FE computations are prohibited. For material models without length scales, once the FE solution of the problem begins to soften in one element, the differential equations of the problem for static codes shift from an elliptic to a hyperbolic system, but the prescribed boundary and initial conditions remain according to an elliptic system of equations, leading to an ill-posed problem (de Borst (1993), Tvergaard and Needleman (1997), and Ramaswamy and Aravas (1998)). The reverse situation exists for dynamics codes. The direct consequence of the loss of ellipticity or hyperbolicity is the appearance of bifurcation with infinite numbers of bifurcated branches (Leblond et al. (1994)), which raises the issue of selecting the relevant one, especially in numerical computations where this drawback manifests itself as a pathological sensitivity of the results with respect to the FE discretization. Also, for problems involving localization, for instance the shear-banding of a 2D rectangular specimen considered in Enakoutsa et al. (2012), localized strain or damage occurs over the smallest element, and the load decreases with decreasing element size. In high-velocity impact damage simulations, adiabatic shear bands due to thermal softening occur (Zukas (1990)); their widths tend to reduce to zero with decreasing element sizes, leading to a spurious material failure without energy dissipation.

The pathological mesh size effects in high-velocity impact numerical simulations have been demonstrated by Borvik et al. (2001a) and Borvik et al. (2001b) for impact failure by ductile damage and plugging, respectively. Bazant et al. (2000) also checked the mesh-size effects in the numerical simulation of the perforation of a concrete wall by a missile and found less mesh sensitivity, despite the fact that the wall obeys a material model exhibiting strain softening. They attributed this "apparent anomaly" to the extremely short duration of the impact event and large inertia effects involved in the problem, which consequently delay localization effects beyond the duration of the event. The simplest way to solve the pathological mesh-sensitivity issues for practical FE problems involving softening, as first proposed by Rousselier (1981), is to put a lower limit on the FE size, as suggested by Redanz et al. (1997). This means that the element size is no longer a mere mathematical artifact but acquires some physical meaning. However, this method is not theoretically optimal. A more elaborate solution consists of developing nonlocal constitutive models, in which a characteristic or internal length scale is included, so that the predicted material deformation is controlled by the material microstructures in addition to stresses and strains. In the work of Abu Al-Rub and Kim (2009), a thermo-visco damage material model embedded with a material internal length scale through the introduction of spatial gradients in the evolution equation of the damage was used to simulate the ballistic penetration and perforation of heterogeneous metal matrix composite targets. Their model predicts the ballistic limit velocities for the Weldox 460E high strength steel circular plates of various thicknesses independently of the mesh sizes, and these predictions are in good agreement with Borvik et al. (2003)'s experiments. The next section addresses a similar high-velocity impact problem, but one with the modified nonlocal material model presented in Section 3.

### 4.2. Simulations of high-velocity impact damage with the local and nonlocal BCJ models

The objective of this section is to report the application of the local and nonlocal BCJ models presented in Section 3 to numerical simulations of a high-velocity impact-induced damage problem using LS-DYNA, which is a general-purpose, FE code for analysis of large deformation-dynamics responses of structures based on explicit time integration. We considered the problem of a 6061-T6 aluminum circular disk of thickness 3.2 mm and 57.2 mm diameter impacted by a rigid penetrator with a given initial velocity (see Figure 6). A similar problem was considered in Bammann et al. (1993) but with a local model.

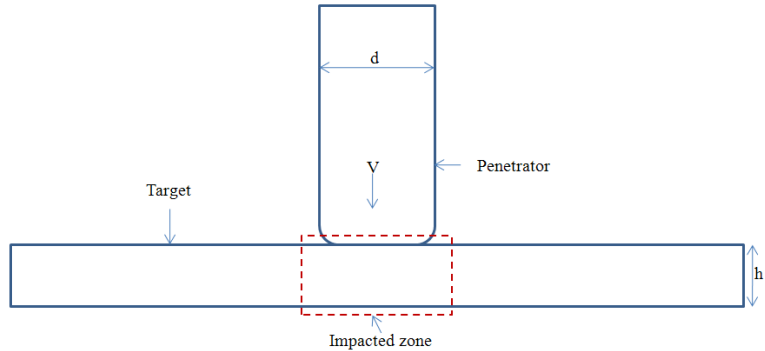

Figure 6: Schematic representation of the initial configuration of the problem model just before the impact.

Four-node quadrilateral axisymmetric shell elements available in the LS-DYNA element library were adopted to represent the disk and the penetrator. The "2D automatic node to surface contact" option was applied with the assumption of frictionless interaction between the rigid penetrator and the disk. Fully-edged, clamped boundary conditions were applied to the disk, while the penetrator was given an initial impact velocity. The disk obeyed both the local and nonlocal BCJ material model behaviors, while the penetrator was assumed to be simply rigid. Extremely small time increments were used to ensure the stability of the numerical computations. The methodology used to determine the BCJ model material parameters for a 6061-T6 aluminum alloy and the parameters are provided in Appendix A. Since the numerical implementation of the BCJ model, based on the Krieg and Krieg (1977)'s radial return method, is already reported in Bammann et al. (1993), we recall the main points of this implementation in Appendix B. The implementation of the integral nonlocal damage Eq. (37) into an existing implicit FE routine is quite involved, since it requires the computation of a double loop over all the integration points of the FE model. However, this difficulty can be removed by calculating the double loop at convergence of global elastic-plastic iterations where the local damage rate at all the integration points as well as the integration point coordinates are known. The advantage of such an algorithm is that it preserves the original architecture of the existing code, while introducing an acceptable approximation: the nonlocal damage term is not computed at each global elastic-plastic iteration but instead once convergence is met. This algorithm was satisfactorily adopted by Enakoutsa et al. (2007) to numerically implement a nonlocal extension of Gurson (1977)'s model. Also, Drabek and Bohm (2005) used the same algorithm in a different context to predict crack paths and overall stress-strain responses that are in good agreement with Baaser and Tvergaard (2003).

The integral nonlocal damage implementation in LS-DYNA, which was used in the numerical simulations below, is based on the previous method. Contrary to the work of Enakoutsa et al. (2007), the convolution operation was performed upon the damage and not its rate. This choice was dictated by the architecture of the code. To verify the choice of convolution operation and calibrate the non-local length scale parameter, we used the notched tensile specimens made of 6061-T6 aluminum alloy to simulate post bifurcation problem and compared the simulated force-displacement curve with available experimental data. Because stress triaxiality is the primary driving factor for damage in porous materials, post-bifurcation issues arise.

In a uniform material, the notch geometry induces a smooth-stress triaxiality field with a maximum value near the center of the specimen. In a damaged medium, stress concentrations induced by the presence of pores may cause local regions of high-stress triaxiality. The notch radii of our specimens were 4 mm and 9.9 mm, respectively. The specimens were 12.7 mm high. A prescribed displacement was applied to both ends of the specimens. For the simulations of the notched tensile problem, only the upper half of the specimen was modeled due to geometric symmetry (see Figure 7). Two different mesh densities were considered for each specimen with a minimum element size of 0.12 and 0.18 mm, respectively. Again, the values of the material parameters for the 6061-T6 aluminum in the absence of the characteristic length scale are provided in Appendix A. Through this calibration process, the characteristic length scale parameter was found to be ~0.5 mm.

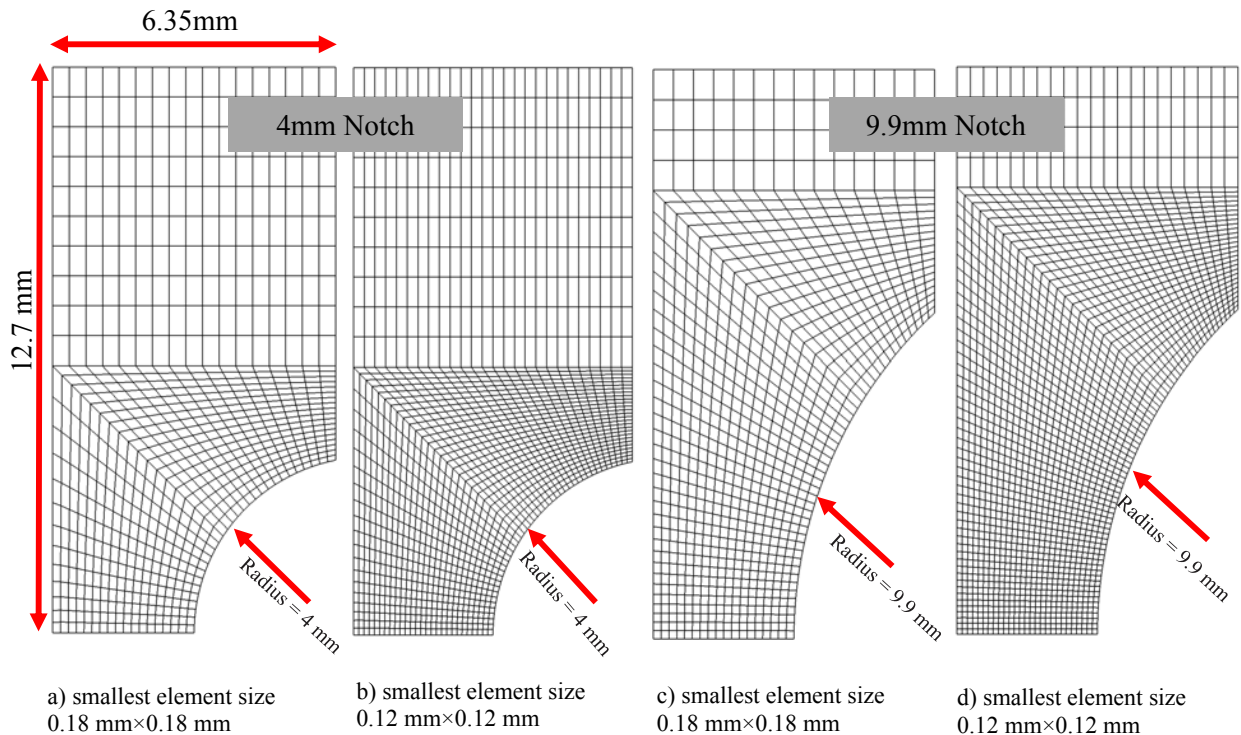

a) smallest element size 0.18 mm×0.18 mm

b) smallest element size 0.12 mm×0.12 mm

c) smallest element size 0.18 mm×0.18 mm

d) smallest element size 0.12 mm×0.12 mm

Figure 7: FE meshes of the notched specimen (a) 4mm radius notch with the minimum element size of 0.18 mm×0.18 mm, (b) 4mm radius notch with the minimum element size of 0.12 mm×0.12 mm, (c) 9.9mm radius notch with the minimum element size of 0.18 mm×0.18 mm, and (d) 9.9mm radius notch with the minimum element size of 0.12 mm×0.12 mm.

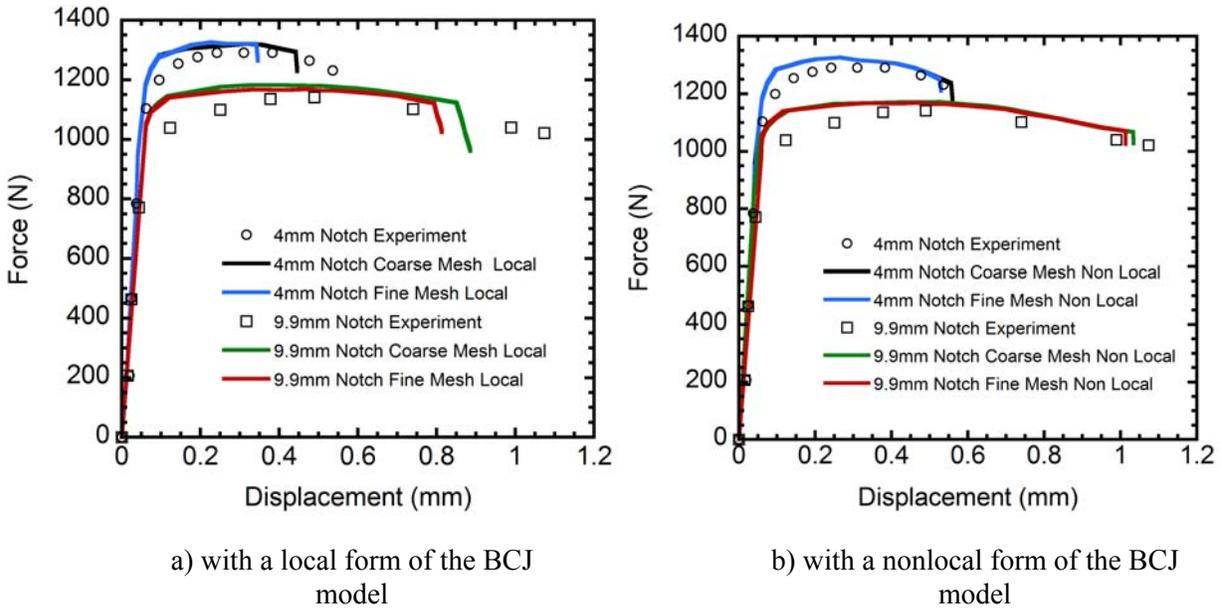

| a) with a local form of the BCJ model | b) with a nonlocal form of the BCJ model |

Figure 8: a) Experimental and predicted force vs. displacement curves for the two mesh densities and the 4 and 9 mm notched radius specimens with the local BCJ model. Note the discrepancies between the curves predicted by the local BCJ model for the two notched specimens. Note also that the rupture of the specimen occurs quickly for the refined mesh in the two specimens. b) Experimental and predicted force vs. displacement curves for the two mesh densities and the 4 and 9 mm notched radius specimens with the nonlocal BCJ model. Note: that the mesh dependency on the predicted load-displacement curves was considerably (if not completely) reduced, when the nonlocal BCJ model was used. Note also that the nonlocal model predicts a delay of the occurrence of the crack in the specimen. The characteristic length scale was found to be equal to 0.5 mm.

Comparison of experimental measured load-displacement responses (Bammann et al. (1996)) of an aluminum alloy 6061-T6 with a local and nonlocal form of BCJ model is shown in Figure 8. Here we used two different notch root radii and two different mesh sizes to compare experimental versus simulated force-displacement curves using the local and nonlocal BCJ models described in Section 3. In the case of the local model, there is a notable effect of the mesh size materialized by an important discrepancy between the two force-displacement curves (see Figure 8a) for the two notched specimens. Furthermore, this discrepancy (mesh dependency) is reduced when the nonlocal BCJ model was used (see Figure 8b). Also, as shown in Figure 8a, the local BCJ model predicted an earlier occurrence of the final failure, contrary to the experimental results, while the occurrence of the final failure predicted by the nonlocal BCJ model was delayed, which compared reasonably well with the experimental results. Thus, the performance of the nonlocal BCJ model to eliminate mesh size effects and predict the fracture process of laboratory-scale tests is verified. Figure 8 shows good correlation and how well the model captured the differences between the stress triaxiality, which expresses the importance of a physically motivated internal state variable plasticity and nonlocal damage continuum model for the predictive modeling.

After good initial correlations with notched specimens, the fracture process of the local versus nonlocal BCJ models under a more complex problem of high-velocity impact damage was compared. Here, we used four different mesh element sizes, as shown in Figure 9, in the impact zone, namely, 0.24, 0.18, 0.12, and 0.08 mm, respectively. To show the effects of the nonlocal concept, i.e. mesh independent solutions, two types of calculations were performed: one with the local BCJ model and the other one with nonlocal form under two different velocities, namely, 89m/s, and 107m/s. For these high-velocity simulations, we used the same characteristic length scale as it was found through calibration of notch specimens.

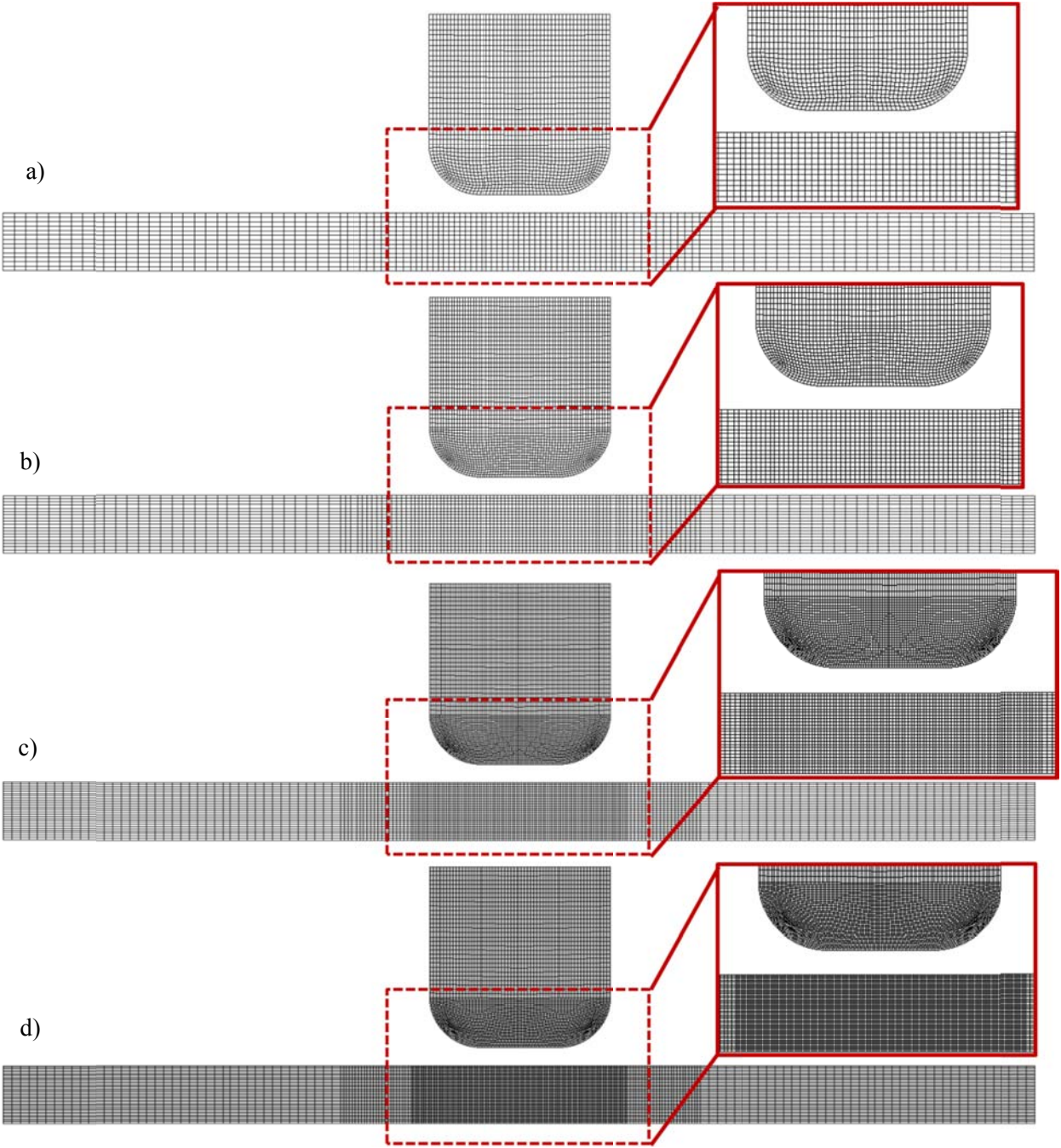

Figure 9: FE meshes of the projectile and the target before the impact, coarse mesh (a), medium mesh (b), fine mesh (c), and refined mesh (d). A uniform mesh is used in the plate domain along the path of the projectile, and the rest of the domain is meshed with gradually coarser mesh. The minimum element size in the impact zone is about 0.24 mm (a), 0.18 mm (b), 0.12 mm (c), and 0.08 mm (d).

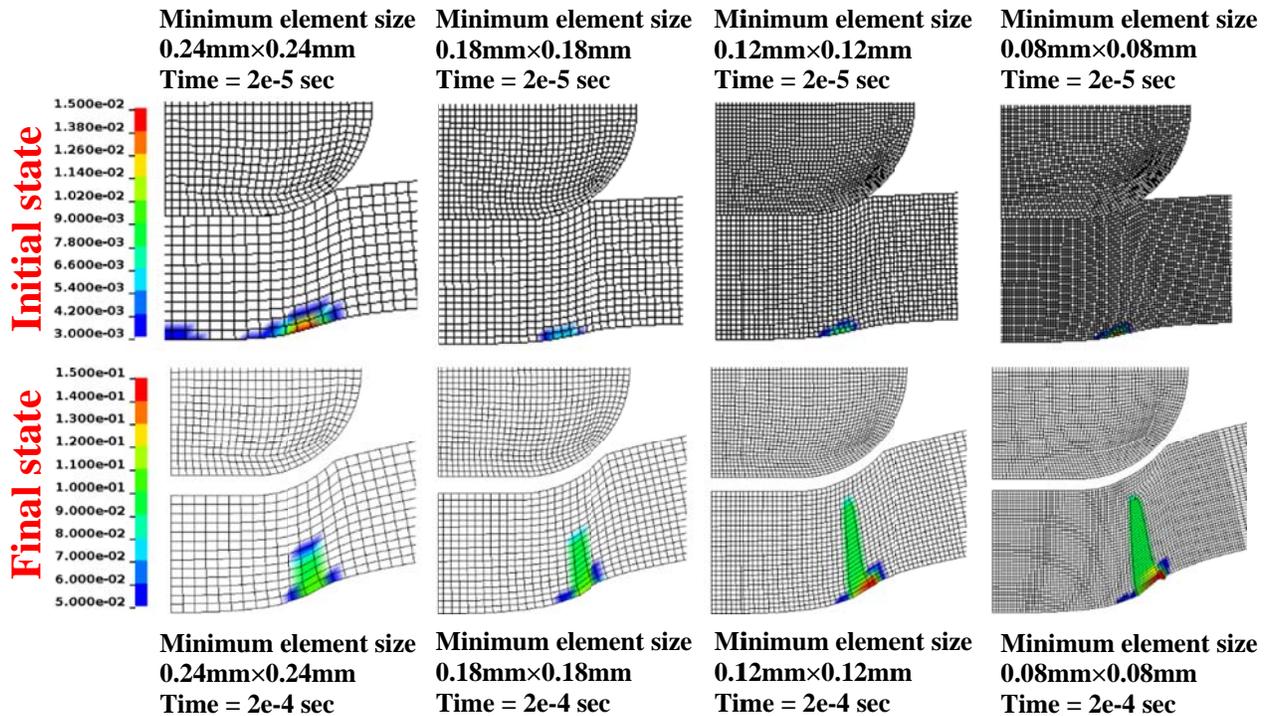

Figure 10: The damage contour plots at the initiation of failure and at the rebounding of the penetrator for four different mesh sizes with an initial impact velocity of 89 m/s. The simulations were carried out using the local BCJ model. We observed that the damage initiates from the back surface of the disk and propagates through its thickness; however, it does not reach the top side. The propagation took place within a layer of one element size, and the length of the damage zone changes as a function of the element size. These two points illustrate the pathological mesh-size effects due to localizations in numerical results.

With the local BCJ model and an applied velocity of 89 m/s, the damage contours at the initiation of failure and at the rebounding of the penetrator for four different mesh sizes are shown in Figure 10. These figures show that the damage concentrates on the back surface of the disk in a single-layer zone that shrinks with decreasing element sizes at the initiation of failure. Note that the peak value/initiation of the damage occurs on the back surface of the disk. The projectile hits the target and rebounds after the failure has initiated. Here also, the damage initiates and propagates in a layer of one element size (see Figure 10), starting from the face of the opposite side of the disk to the impact zone with no clear plug formation. In the case of 107 m/s applied velocity with the local BCJ model, we observed similar behavior of damage localizations when compared with the 89 m/s velocity case, i.e., the damage concentrates on the back surface of the disk in a single-layer zone that shrinks with decreasing element sizes. However, contrary to the simulations with an initial impact velocity equal to 89 m/s, once the failed zone reached the top side of the disk, a plug was fully cut from the disk through the one element size damaged layer. As we would expect with the local form of the BCJ model, the pathological mesh dependency was clearly visible. With an increase in the mesh refinement, the damage zone size increased significantly. From the protective-system design point of view, the pathological mesh size effects may over or under-estimate the predicted material characteristics, which include but are not limited to the impact absorption capacity and the ballistic limit, required to resist a given impact-loading condition. The improvements brought by the introduction of the nonlocal concept in the BCJ model are outlined below.

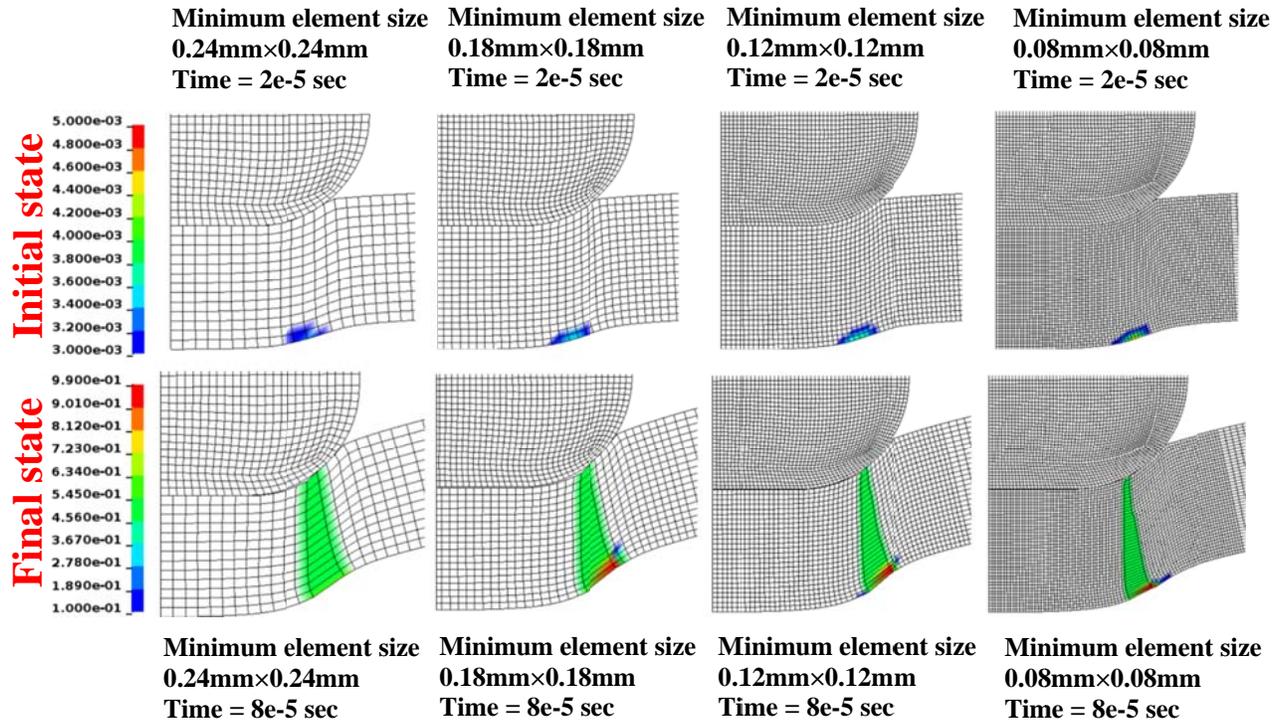

Figure 11: The damage contour plots at the initiation of failure and at the maximum penetration of the penetrator for four different mesh sizes with an initial impact velocity of 107 m/s. The simulations were carried out using the local BCJ model. We observed that the damage initiates from the back surface of the disk and propagates through its thickness to the top side. The propagation took place within a layer of one element size and the length of the damage zone changes as function of the element size. These two points illustrate the pathological mesh-size effects due to localizations in numerical results.

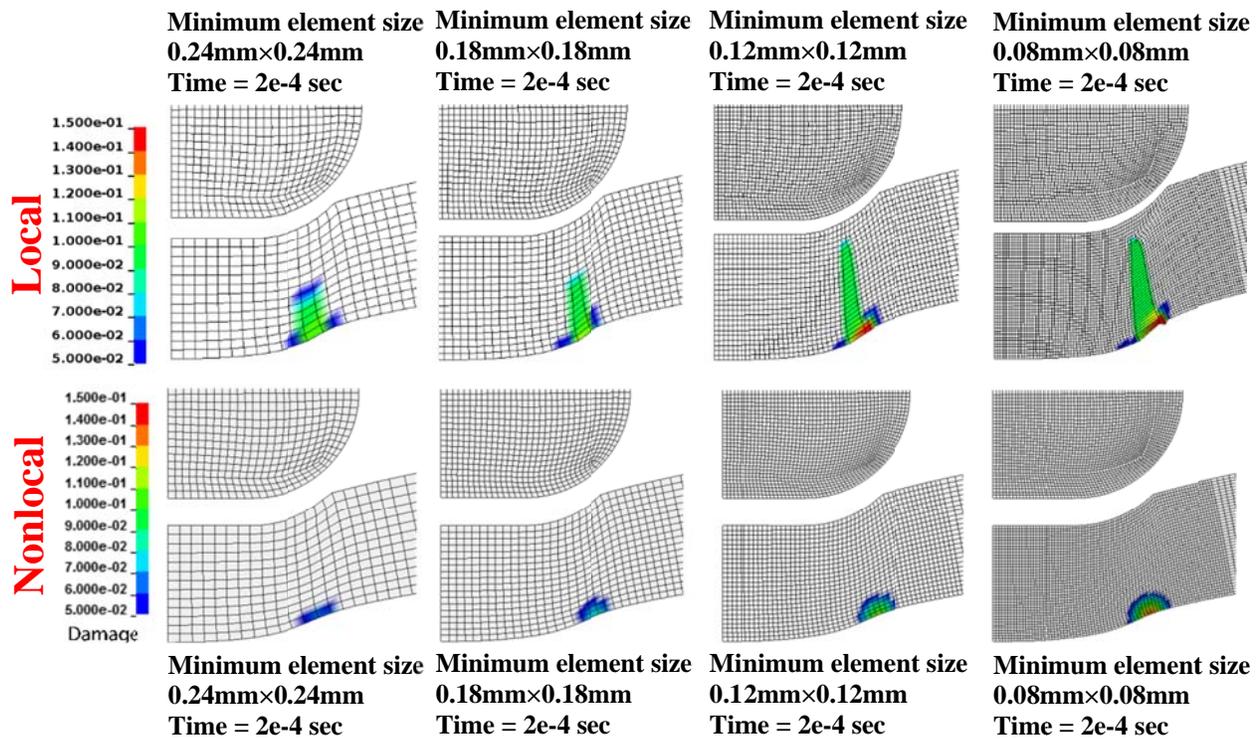

Figure 12: The damage contour plot comparisons at the rebounding of the penetrator for four different mesh sizes simulated using the local and nonlocal BCJ model with an initial impact velocity of 89 m/s. In the case of the nonlocal model, the localization of the damage still take place, however, the localization zone does not shrink with the mesh refinement as compared to the local model. Also, in the case of simulations with the nonlocal model we observed approximately same size damage zone indicating mesh independent numerical solutions.

The comparisons of damage behaviors under an initial impact velocity of 89 m/s for the local and nonlocal models are shown in Figure 12. In the case of the nonlocal model, the localization of the damage still occurred after the initiation of the damage on the back surface of the disk. However, the localization zone did not shrink with the mesh refinement as compared to the local model. More specifically, the damage zone spread over several elements. Furthermore, it is interesting to point out the quasi-similarity of the damage pattern for the four mesh discretizations in contrast to the damage patterns for the local BCJ model. The quasi-similarity of the damage pattern, i.e., the damage zone size predicted by all four meshes, has same order of magnitudes, suggesting a mesh-independent numerical solution.

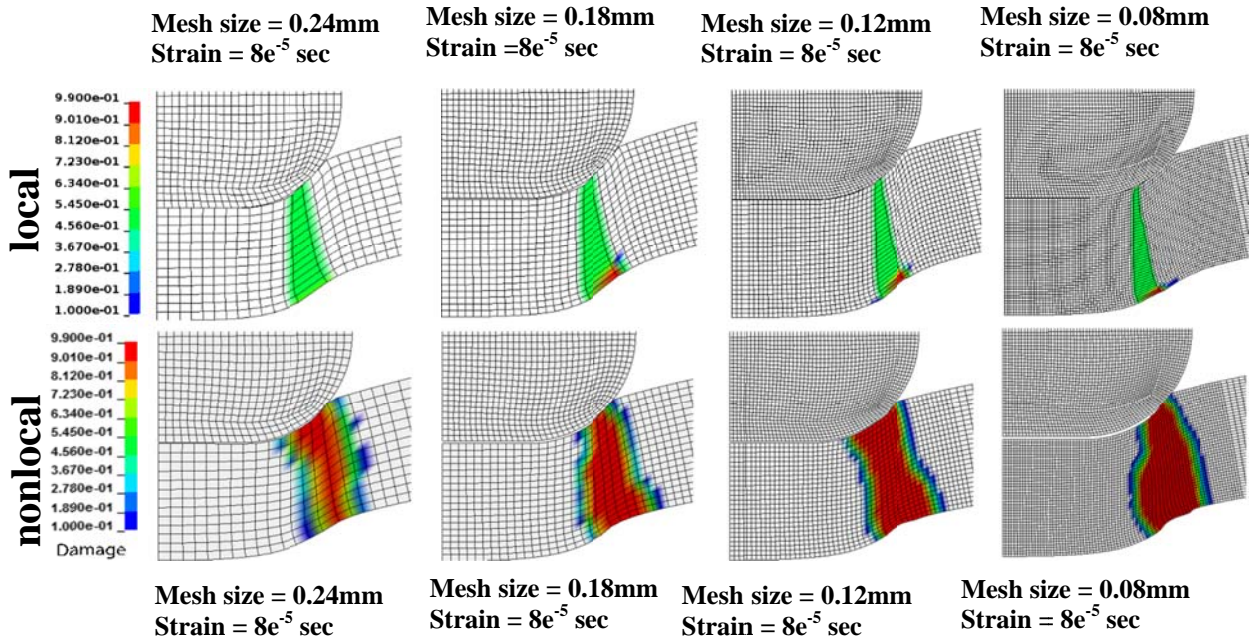

Figure 13: The damage contour plot comparisons at the maximum penetration of the penetrator for four different mesh sizes simulated using the local and nonlocal BCJ models with an initial impact velocity of 107 m/s. In the case of the nonlocal model, the localization of the damage still take place, however, the localization zone does not shrink with the mesh refinement as compared to the local model. Also, in the case of simulations with the nonlocal model we observed approximately same size damage zone indicating mesh independent numerical solutions. The influence of the damage delocalization is clearly demonstrated: the damage propagates within layer of width more than one element, the damage zone is quasi-similar is the four cases considered.

In the case of initial applied velocity of 107 m/s, damage behaviors of various mesh using the local and the nonlocal BCJ models is shown in Figure 13. Here, we observed similar behavior when compared with the nonlocal BCJ model with the initial velocity of 89 m/s, i.e., the damage zone did not shrink with the mesh refinement. Also, in the case of simulations with the nonlocal model, we observed approximately the same damage-zone width, indicating mesh-independent numerical solutions. Several other points of interest can be observed from Figure 13. The first one is that the damage zone predicted by the use of the nonlocal BCJ model gives way to a more smoothed zone wherein a sharp damage gradient no longer exists. Furthermore, the failure initiation predicted by the nonlocal BCJ model was delayed with respect to the corresponding one predicted by the local version of the BCJ model. The plausible explanation of this phenomenon is that the convolution integral of the damage does not allow the damage to reach rapidly high values, so that the value of the damage at a given material point (in the impact region, more specifically) reaches the critical failure value[4] at a later times.

---

[4] In the BCJ model, failure is obtained at a material point once the damage in that point reaches a critical value.

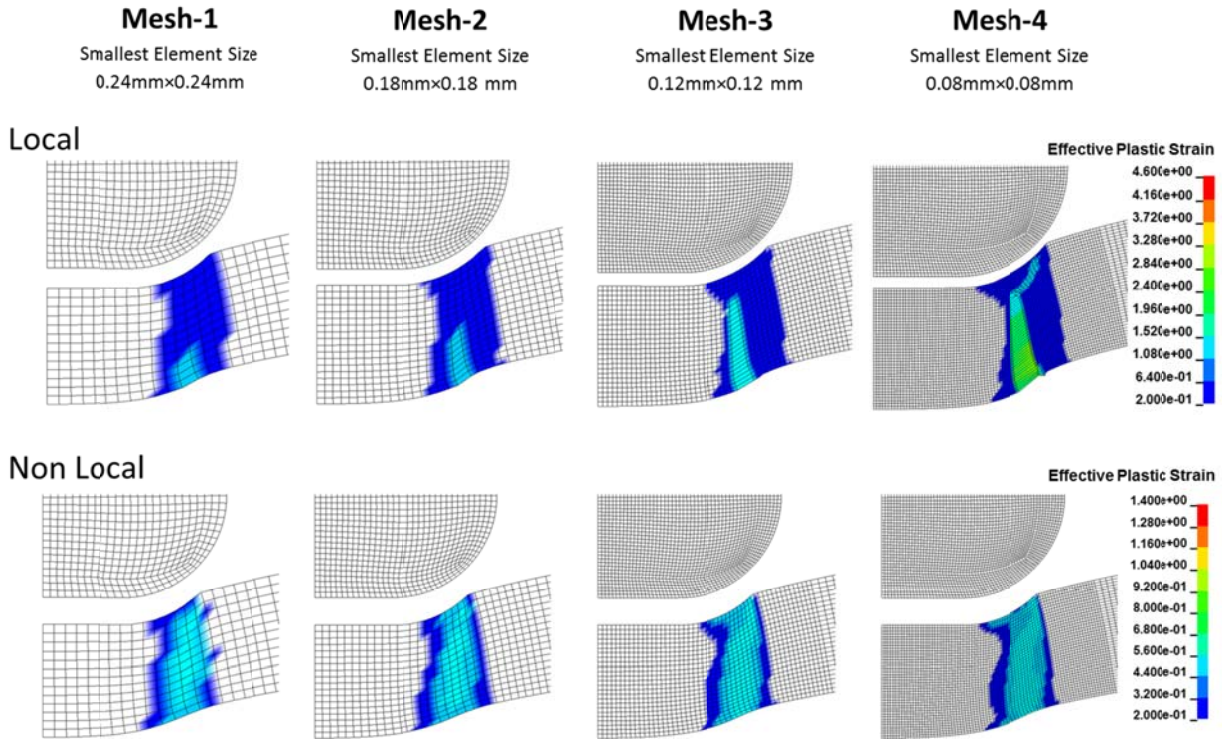

Figure 14: The effective plastic strain contour plot comparisons at t= 2×10⁻⁴ sec for four different mesh sizes simulated using the local and nonlocal BCJ models with an initial impact velocity of 89 m/s. Interestingly, the regularization effect of the non-local enhancement becomes clear with the contour plots of effective plastic strain. Furthermore, the effective plastic zone/band are smoother with the nonlocal BCJ model when compare with the local model.

The comparisons of effective plastic strain behaviors under an initial impact velocity of 89 m/s and 107 m/s for the local and nonlocal models are shown in Figures 14 and 15 respectively. The regularization effect of the nonlocal enhancement becomes clear with the contour plots of effective plastic strain. The effective plastic zone/band computed through the nonlocal model are more smoothed as the effective plastic strains spread out into the neighboring elastic region, i.e., sharp gradient no longer exists. On the other hand, the effective plastic strain band/zone computed through the local model reveals the existence of strong localization into a line, i.e., no broadening of the localization zone, which emanates from the lower-bottom of the plate. Furthermore, the effective plastic strain computed through the local model shows significant mesh dependency, i.e., the size of the localized zone changes significantly with reduction in mesh size, i.e., the pathological mesh-size effects. Similar behavior was observed in the case with the initial impact velocity of 107 m/s, see Figure 15.

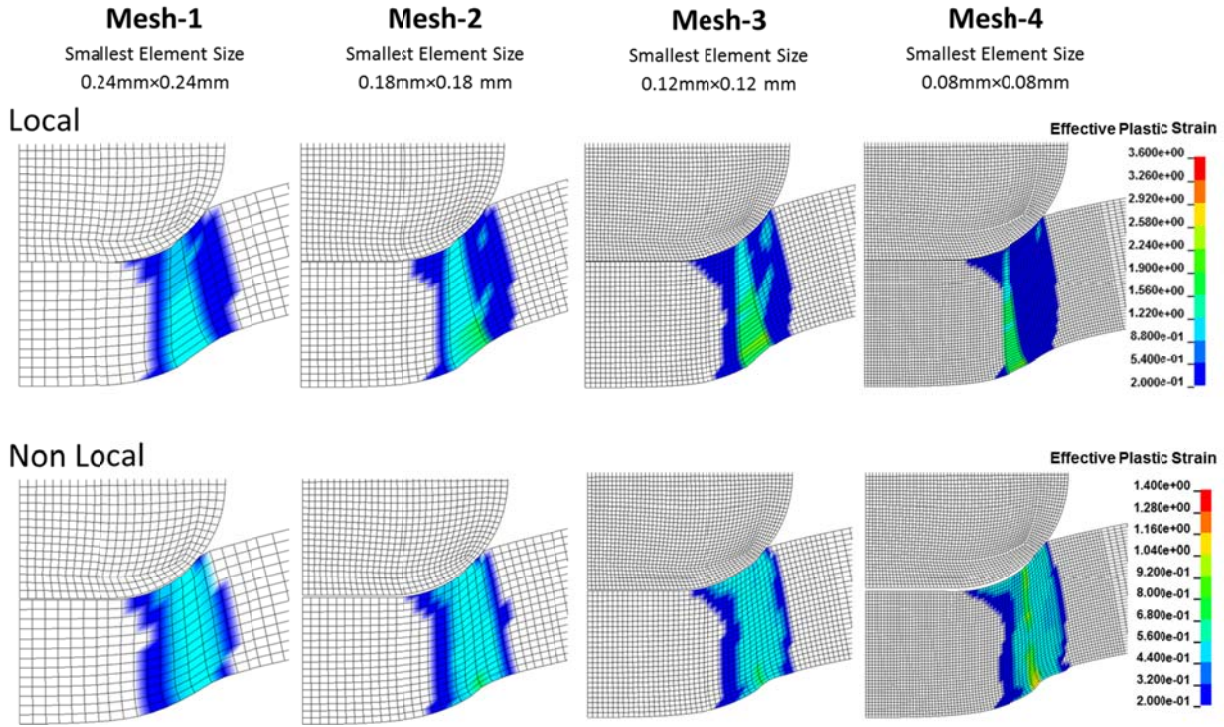

Figure 15: The effective plastic strain contour plot comparisons at t= 8×10⁻⁵ sec for four different mesh sizes simulated using the local and nonlocal BCJ models with an initial impact velocity of 107 m/s. Interestingly, the regularization effect of the non-local enhancement becomes clear with the contour plots of effective plastic strain. Furthermore, the effective plastic zone/band are smoother with the nonlocal BCJ model when compare with the local model.

It remains the ultimate question of whether the nonlocal BCJ model predictions are comparable to the experimental observed results of high-velocity impact tests. To address this question, we considered the failure in aluminum disks impacted by steel bars in experiments performed at Sandia National Laboratory and reported in Bammann et al. (1993). In these experiments, 6061-T6 aluminum disks (57.2 mm diameter; 3.2 mm thick) were impacted with a hardened steel road. The disks were held in a manner that simulated free boundary conditions, similar to the modeling constrained applied here in this study. The results of these experiments show that initial failure, in the form of visible cracks on the side of the disk opposite the bar, occurred at an impact velocity between 79 to 84 m/s (Figure 16), and the disk failed completely through its thickness at an impact velocity between 92 to 107 m/s (Figure 17). The nonlocal BCJ model, with a characteristic length scale calibrated through a series of tensile notch tests and an impact velocity of 107 m/s, predicts that a plug is fully cut from the disk, and this exactly corresponds to the experimental results (see Figure 17). With the same characteristic length scale and at a lower velocity of 89 m/s, predicts a crack initiates at the back side of the plate but does not reach the other side of the disk (see Figure 16). Figures 16 and 17 demonstrate the satisfactory comparison between predicted failure and the experimental results reported in Bammann et al. (1993).

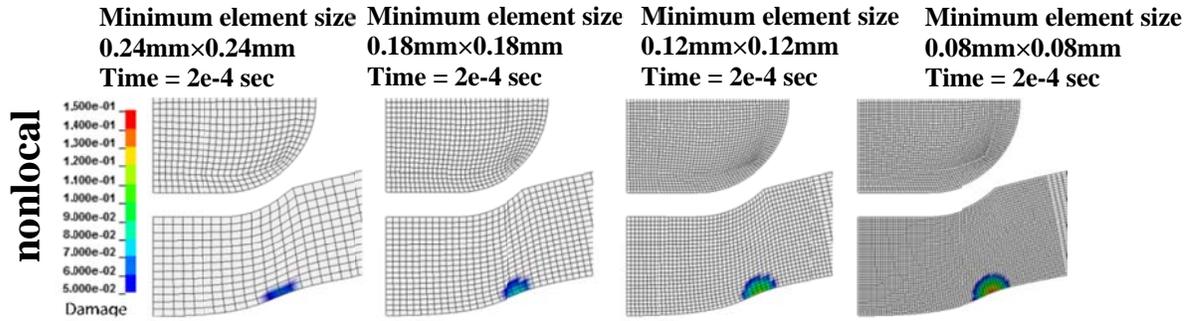

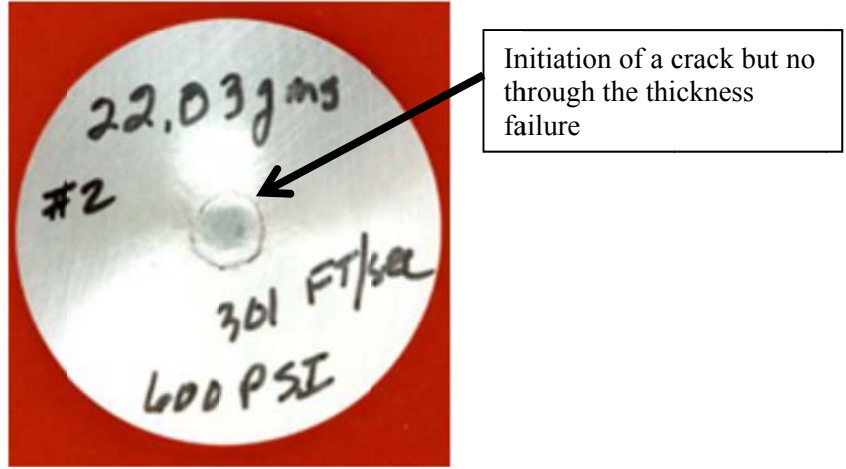

Figure 16: Comparison of Sandia National Laboratory experimental results (in Bammann et al. (1993)) with the numerical simulations using the nonlocal BCJ model for an impact velocity of 89 m/s. A crack is initiated on the convex side of the disk, but it does not reach the opposite side of the disk and show an excellent agreement with the mesh independent FE results.

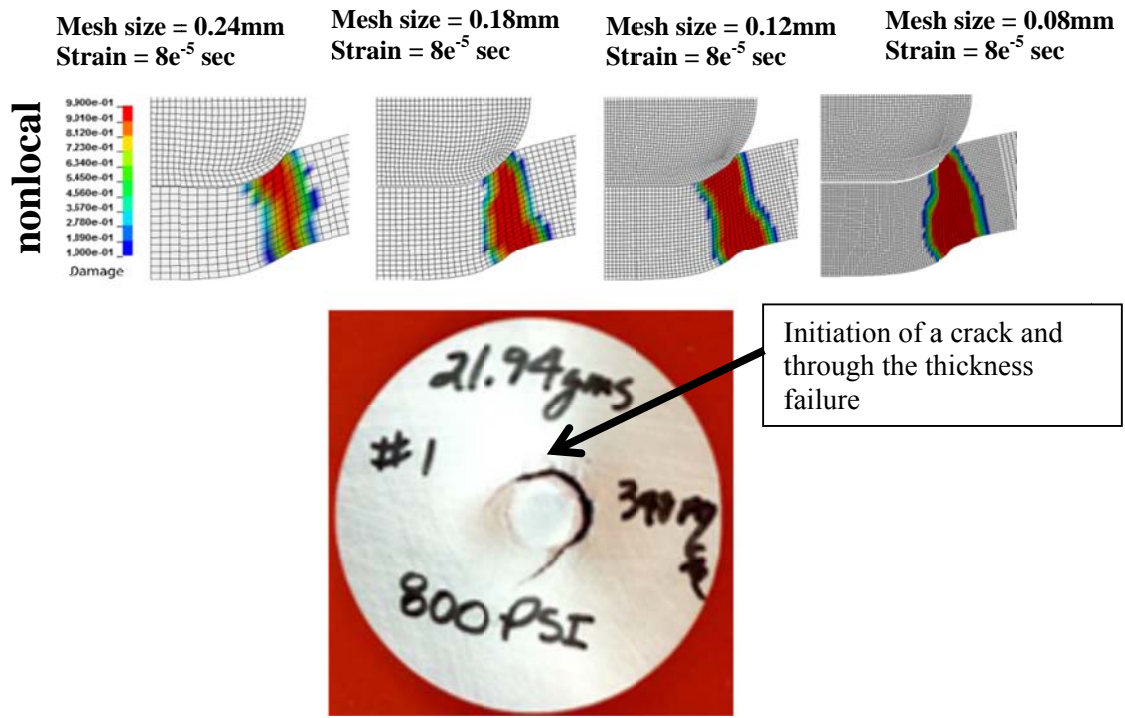

Figure 17: Comparison of Sandia National Laboratory experimental results (in Bammann et al. (1993)) with the FE analyses using the nonlocal BCJ model for an impact velocity of 107 m/s. A crack was observed through the full thickness of the disk and shows an excellent agreement with the mesh independent FE results.

## 5. Summary and perspectives

This paper was devoted to (1) predicting mesh-independent FE simulations of high-velocity impact of a rigid projectile colliding against a 6061-T6 aluminum disk described by the BCJ material model but modified with a nonlocal evolution equation for the damage through the manner proposed by Pijaudier-Cabot and Bazant (1987) and (2) comparing model predictions with experimental data. The motivation was to propose a reliable analytical and computational tool that can be utilized to increase the performance of material under high-velocity impact loading conditions and address the issue of calibration of the nonlocal length scale parameter. Two examples comparing the predictions of the modified BCJ model with experimental data were presented. These include modeling the failure of a series of notched-specimens of 6061-T6 aluminum to calibrate the nonlocal length scale parameter and the penetration of an aluminum disk by a projectile impacting at different velocities. The results of the numerical simulations show that the addition of a characteristic length scale to the local BCJ model does eliminate the pathological mesh-size effects obtained in the practical applications of the classical BCJ model. In addition the numerical predictions of the new model are in good agreement with experimental data. This establishes the robustness of the delocalization method to deal with the failure process in the numerical simulations of problems involving high-rate deformation and damage of aluminum structures. The assessment of the damage delocalization technique for other metallic materials such as steels and titanium under high-rate deformation will be investigated in the future. Also, future work will include the significance of the length scale introduced in the BCJ model by the nonlocal concept in modeling high-rate deformation of metallic materials. What is certain from this study is that the length scale introduced in the model can be calibrated using a series of tensile notch specimen tests, however, it is not clear how they are related to the physical and constitutive features of the material response under the impact loading.

**Acknowledgments**

The study was supported by the U.S. Department of Transportation, Office of the Secretary, Grant No. DTOS59-08-G-00103. Dr. Lee Binderman from LSTC Inc. is gratefully thanked for his insightful discussions on the use of the nonlocal concept in LS-DYNA FE code. KNS would like to acknowledge support from the School for Engineering of Matter, Transport, and Energy (SEMTE) at Arizona State University.

## Appendix A. Determination of the BCJ material parameters for 6061-T6 Aluminum

This appendix presents the method used to determine the BCJ model parameters for 6061-T6 aluminum. We shall distinguish two set of parameters: the plasticity parameters and the damage parameters. The procedure to determine the plasticity parameters is extensively discussed in Bammann (1990a) and is based on the uniaxial stress at constant true strain rate and temperature. Since the plastic strain rate can be reasonably approximated by the total strain rate at large strain (viscoplastic assumption), the plastic strain rate is substituted by the total strain rate. Then the flow rule is inverted giving an expression similar to the yield surface, and the equations of the kinematic and isotropic variables are analytically integrated to obtain the closed form solutions, i.e.,

$$\alpha = \sqrt{\frac{h\dot{\varepsilon}}{r_d\dot{\varepsilon}+r_s}}\tanh\left[\sqrt{\frac{h(r_d\dot{\varepsilon}+r_s)}{\dot{\varepsilon}}}\varepsilon\right] \tag{A1}$$

$$\kappa = \sqrt{\frac{H\dot{\varepsilon}}{R_d\dot{\varepsilon}+R_s}}\tanh\left[\sqrt{\frac{h(R_d\dot{\varepsilon}+R_s)}{\dot{\varepsilon}}}\varepsilon\right] \tag{A2}$$

$$\sigma = \alpha + \left(\kappa + Y(\theta)\right) + V(\theta)\sinh^{-1}\left[\frac{\|\dot{\varepsilon}\|}{f(\theta)}\right] \tag{A3}$$

where, $\sigma$ and $\varepsilon$ represent the true stress and strain in a uniaxial tension or compression, and $\alpha$ is the uniaxial component of the tensor $\boldsymbol{\alpha}$.

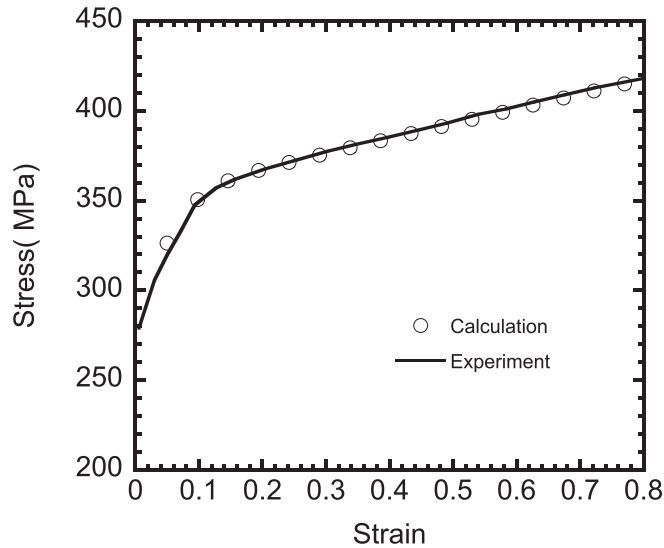

Figure A1: Experimental and analytical stress strain responses of Al 6061-T6 for room temperature and at a constant strain rate.

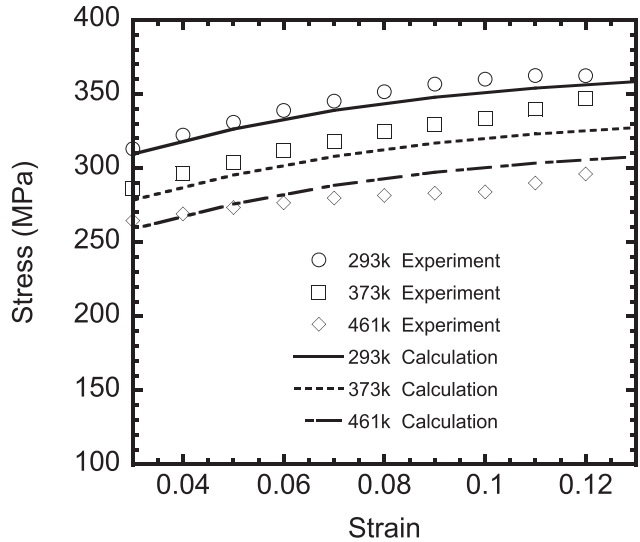

Figure A2: Experimental and simulated stress strain responses of Al 6061-T6 at three different temperatures for a constant strain rate.

The inelasticity parameters for the 6061-T6 aluminum are identified using the experimental data from Bammann et al. (1990). One uniaxial large strain compression test at 293K (see Figure A1) and three uniaxial tension tests carried out at 293K, 373K, and 461K (see Figure A2) are used to obtain the inelastic parameters. The damage model parameters (i.e., initial damage volume fraction, damage growth parameter, and characteristic length-scale) for T6-6061 aluminum are determined using the notch tensile test data taken from Bammann et al. (1996), see Figure 8. Because of the significant tensile pressure generation, the notch radius dominates the damage evolution in metal (Bammann et al., 1990b), and notch specimens of different radii are typically used to determine damage parameters. Nevertheless, the notch tensile test is a three dimensional boundary value problem with no closed form solution. Thus, a trial and error finite element analysis of axisymmetric notch tensile specimens is utilized to identify the damage parameters. In this study, the initial damage volume fraction is assumed and the rest of the damage parameters are obtained from the 4mm radius notch test. Next, the 9.9mm radius notch test is used for the model verification. The damage parameters in addition to the inelastic constants found using the methodology described above are summarized below.

Table A1. The model parameters for 6061-T6 aluminum alloys

| Parameters | Unit | Value |
|---|---|---|
| C1 | Mpa | 0.00 |
| C2 | K | 1.00 |
| C3 | Mpa | 160.00 |
| C4 | K | 161.70 |
| C5 | 1/sec | 1.00 |
| C6 | K | 0.00 |
| C7 | 1/Mpa | 0.194 |
| C8 | K | 694.40 |
| C9 | Mpa | 1027.32 |
| C10 | K | 0.00 |
| C11 | 1/(Mpa-sec) | 0.00 |
| C12 | K | 0.00 |
| C13 | 1/Mpa | 0.00439 |
| C14 | K | 855.560 |
| C15 | Mpa | 83.00 |
| C16 | K | 0.00 |
| C17 | 1/(Mpa-sec) | 0.00 |
| C18 | K | 0.00 |
| $\phi_0$ | - | 0.001 |
| m | - | 8 |

**Appendix B. Main elements of the numerical implementation of the BCJ model**

The objective of this appendix is to present the numerical implementation of the BCJ model in a finite element code. This implementation is based on (i) the use of Krieg and Krieg (1977)'s radial return proposal to solve the equations presented in Section 2 and (ii) the formulation of a numerical flow rule at each time step derived from a numerical consistency condition. The implementation of the model into LS-DYNA is described following the flowchart presented below.

- First, the temperature and rate functions are evaluated using $\theta^N$ and $\boldsymbol{D}_p^{N+1}$, where the N superscript denotes the value at time step N, and N+1 denotes the value at time step $N+1$.

- Then, the trial stresses and state variables are calculated assuming elasticity following the formula:

$$\boldsymbol{\sigma}^{TR} = \lambda(1-\phi^N)tr(\boldsymbol{D}_p^{N+1})\boldsymbol{I} + 2\mu(1-\phi^N)\boldsymbol{D}_p^{N+1} - \frac{\dot{\phi}^N}{1-\phi^N}\boldsymbol{\sigma_N} \tag{B1}$$

$$p^{N+1} = \frac{1}{3}tr\boldsymbol{\sigma}^{TR}, \qquad \boldsymbol{\sigma}^{TR} = \boldsymbol{\sigma}^{TR} - p^{N+1}\boldsymbol{I} \tag{B2}$$

$$\boldsymbol{\alpha^{TR}} = \boldsymbol{\alpha^N} - \left(r_s + r_d\sqrt{\frac{2}{3}}\|\boldsymbol{D}_p^{N+1}\|\right)\|\boldsymbol{\alpha}\|\boldsymbol{\alpha} \tag{B3}$$

$$\kappa^{TR} = \kappa^N - \left(R_s + R_d\sqrt{\frac{2}{3}}\|\boldsymbol{D}_p^{N+1}\|\right)\kappa^2 \tag{B4}$$

- The elastic assumption is now checked by substitution into the yield condition

$$\chi^{TR} = \left\|\boldsymbol{\sigma}^{TR} - \frac{1}{3}\boldsymbol{\alpha^{TR}}\right\| - \sqrt{\frac{2}{3}}(\kappa^{TR} + Y(\theta))(1-\phi^N) \tag{B5}$$

- If $(\chi^{TR}) \le 0$ then $\boldsymbol{\sigma} = \boldsymbol{\sigma^{TR}}, \quad \boldsymbol{\alpha} = \boldsymbol{\alpha^{TR}}, \quad \kappa = \kappa^{TR}$, and the other internal state variables and their associated variable remain constant:

$$\boldsymbol{\varepsilon_p} = (\boldsymbol{\varepsilon_p})^N, \text{ and } \phi = \phi^N \tag{B6}$$

In the case that the stress does not lie within the yield surface (B5) we have to apply the plastic corrector procedure on the yield surface. If $(\chi^{TR}) > 0$ then the plastic deformation occurred and the trial stresses must be returned to the new yield surface:

$$\boldsymbol{\sigma}'_{N+1} = \boldsymbol{\sigma'^{TR}} - \int 2\mu\boldsymbol{D_p}dt$$
$$\boldsymbol{\alpha}'_{N+1} = \boldsymbol{\alpha'^{TR}} + \int h\boldsymbol{D_p}dt$$
$$\kappa'_{N+1} = \kappa'^{TR} + \int \sqrt{\frac{2}{3}}H\|\boldsymbol{D}\|dt \tag{B7}$$

- Using the radial return method, the assumption is then made that the plastic strain rate is constant over the time step and in the direction of the effective stress:

$$\boldsymbol{\zeta} = \boldsymbol{\sigma}^{TR} - \frac{2}{3}\mu\boldsymbol{\alpha}^{TR}$$

$$\boldsymbol{n} = \frac{\boldsymbol{\zeta}}{\|\boldsymbol{\zeta}\|}$$

$$\int \boldsymbol{D_p}\,dt = \Delta\gamma\,\boldsymbol{n} \tag{B8}$$

- Substituting the expression for the plastic strain increment yields

$$\boldsymbol{\sigma}'_{N+1} = \boldsymbol{\sigma}'^{TR} - 2\mu\Delta\gamma\,\boldsymbol{n}$$

$$\boldsymbol{\alpha}'_{N+1} = \boldsymbol{\alpha}'^{TR} + h\Delta\gamma\,\boldsymbol{n}$$

$$\kappa'_{N+1} = \kappa'^{TR} + \sqrt{\frac{2}{3}}H2\mu\Delta\gamma \tag{B9}$$

- These expressions are then used in the consistency condition ($\chi = 0$) to solve for $\Delta\gamma$. (Note that for the kinematic hardening this is not a straightforward solution.) For the assumption made here, this result in a linear algebraic equation. It is then a simple matter to find the stress and the state variables at t=N+1 by substituting the value for $\Delta\gamma$ into Eq.(B9):

$$\Delta\gamma = \frac{\chi^{TR}}{2\mu + \frac{2}{3}(h+H)} \tag{B10}$$

- Temperature is then updated and damage is updated assuming stresses were constant over the time step. The equation rate for void growth can be very stiff numerically, and integration using an Euler forward difference can generate numerical errors. The evolution equation of the damage, Eq. (22), however, can be integrated analytically for constant values of $p$, $\bar{\sigma}$, and $\varepsilon_p$. Note that for 32-bit computers the "exact" solution can contain numerical round off errors, so the Euler method must be used. The exact solution is used on 64-bit machines or 32-bit machines with double precision. The exact solution is

$$\varsigma = \sinh\left[\frac{2(2m-1)}{(2m-1)}\frac{p}{\{\|\boldsymbol{\sigma}'-\boldsymbol{\alpha}\|\}}\right] \tag{B11}$$

$$\phi^{N+1} = 1 - \left[1 + \{(1-\phi^N)^{1+m} - 1\}\exp\langle\|\boldsymbol{D_p}\|\varsigma(1+m)\Delta t\rangle\right]^{1/1+m} \tag{B12}$$

- The Euler method solution is

$$\phi^{N+1} = \phi^N + \Delta\dot{\phi}^{N+1} \tag{B13}$$